 \documentclass[letterpage,11pt]{article}
 \usepackage[english]{babel}
 \usepackage[pdftex]{graphicx}
  \usepackage{wrapfig,fullpage}
  \usepackage{fancybox,enumerate, floatrow}
  \usepackage{setspace}
  \usepackage{fancyvrb}
  \usepackage{amsfonts,amsmath,amsthm,array,amssymb}
\usepackage{hyperref}
\usepackage{color,xcolor}
\usepackage{lscape}
\usepackage{mathtools}

\usepackage{verbatim}

 \include{latexsymb}
 \include{pxfonts}
 \include{displaymath}
 \include{tipa}

\begin{document}

\begin{center}

{\bf  Bivariate binomial conditionals distributions with positive and negative correlations: A statistical study}
\end{center}

\bigskip
\begin{center}
{  Indranil Ghosh$^{1},$ Filipe Marques$^{2},$ Subrata Chakraborty$^{3}$\\
{\small $^{1}${University of North Carolina, Wilmington, USA}\\
 $^{2}${Universidade  Nova de Lisboa, Portugal}\\
  $^{3}${Dibrugarh University, Assam, India}\\
  Corresponding author email address: ghoshi@uncw.edu
}}
\end{center}

\bigskip

\begin{abstract}
In this article, we discuss a bivariate  distribution whose conditionals are  univariate  binomial distributions and the marginals are not binomial that exhibits negative correlation. Some useful structural properties of this distribution namely marginals, moments, generating functions, stochastic ordering are investigated. Simple proofs of negative correlation, marginal over-dispersion, distribution of sum and conditional given the sum are also derived. The distribution is shown to be a member of the multi-parameter exponential family and some natural but useful consequences are also outlined.  The proposed distribution tends to a recently investigated conditional Poisson distribution  studied by Ghosh et al. (2020). Finally, the distribution is fitted to two bivariate count data sets with  an inherent negative correlation  to  illustrate its suitability.
\end{abstract}

\medskip

\noindent {\bf Keywords:} Bivariate binomial distribution, Conditional specification, Negative and positive correlation, Conditional failure rate, Limiting distribution.

\section{Introduction}
The study of correlation between binomial random variables is an important statistical problem with a lot of theoretical and practical applications and is not new in the literature, see Biswas et al. (2022) and the references cited therein. There are also some attempts finding  bivariate binomial distributions in other directions. Hamdan et al. (1971) introduced a bivariate binomial distribution (which is indeed, a bivariate compound Poisson distribution). Hamdan et al. (1976) studied  the joint distribution of the total numbers of occurrences of binary characters $A$ and $B,$ given three independent samples in which both characters, $A$ but not $B,$ and $B$ but not $A,$ are observed---effectively derived a bivariate binomial distribution.
A symmetric bivariate binomial distribution was proposed by Le (1984)  to analyze clustered samples in medical research.  Papageorgiou et al. (1994) examined mixtures of bivariate binomial distributions which were derived from bivariate-compounded Poisson distribution. Ling et al. (1989) discussed bivariate binomial distributions from extension of classes of univariate discrete distributions of order $k.$ Takeuchi et al. (1987) obtained the sum of  $0-1$ random variables in the multivariate setup. A bivariate generalization of the three parameter quasi-binomial distribution of Consul (1979) has been obtained by Mishra et al. (1996). Crowder et al. (1989) carried out Bayesian inference, and they defined the bivariate binomial distribution in a different sense. They defined a two-fold binomial model like $X_1|m \sim Bin\left(m, p\right)$ and $X_{2}|X_{1}; m\sim Bin\left(X_1, q\right).$  For some discussions on bivariate binomial distributions see Kocherlakota et al. (1992) and Johnson et al. (1997).
However, none of the above cited references considered the construction and study of a bivariate distribution such that both the conditionals are binomial with respective parameters, but the joint distribution may not necessarily be a bivariate binomial in its traditional sense. Arnold et al. (1999) came up with the construction of a bivariate discrete distribution starting from two conditional distributions that are binomial. In fact this idea is first developed in Arnold et al. (1991). It appears that the resulting bivariate discrete model (albeit ubiquitous normalizing constant) has the salient feature of exhibiting both positive and negative correlation---this property is not enjoyed by many of the existing bivariate binomial models. In this article, we explore some useful structural properties of the bivariate discrete distribution originally proposed by Arnold et al. (1991,1999) and discusses its applicability in modeling  bivariate discrete  data exhibiting either negative  correlation. The rest of the paper is organized as follows. In Section $2,$ we introduce the bi-variate binomial conditionals distribution  which was described in  Arnold et al. (1991,1999)  and provide the expression for associated marginal p.m.f.'s,  of $X$ and $Y.$ In Section $3,$ we provide several useful structural properties of this distribution.  Section $4$ provides the estimation of parameters using sample proportions and also  under the maximum likelihood method.  In  section $5,$ we discuss copula-based simulation.  A real data application for the BBCD is presented in section $6.$  Finally, some concluding remarks are provided in Section $7.$

\section{Bivariate  binomial conditional distributions}

Let us assume the following: 

\begin{itemize}
\item $X|Y=y\sim \text{binomial}\left(n_{1}, p_{1}(y)\right),$  for each fixed $Y=y.$

\item $Y|X=x\sim \text{binomial}\left(n_{2}, p_{2}(y)\right),$ for each fixed $X=x.$
\end{itemize}

\noindent  According to \cite{Acs, as}, the associated joint p.m.f. will be

\begin{equation}
\label{pmf}
P\left(X=x, Y=y\right)
=K_{B}\left(n_{1}, n_{2},p_{1}, p_{2}, t \right)\binom{n_{1}}{x} \binom{n_{2}}{y} p^{x}_{1} p^{y}_{2} \left(1-p_{1}\right)^{n_{1}-x} \left(1-p_{2}\right)^{n_{2}-y} t^{xy},
\end{equation}

\noindent where $p_{1}\in (0,1),$ and $p_{2}\in (0,1),$ and $t>0,$ and  $x=0,1,2,\cdots, n_{1}; \quad y=0,1,2,\cdots, n_{2};$ and $K_{B}\left(n_{1}, n_{2},p_{1}, p_{2}, t \right)$ is the normalizing constant and

\begin{equation*}
K_B^{-1}=K^{-1}_{B}\left(n_{1}, n_{2},p_{1}, p_{2}, t \right)
=\sum_{x=0}^{n_1}\sum_{y=0}^{n_2}\binom{n_{1}}{x} \binom{n_{2}}{y} p^{x}_{1} p^{y}_{2} \left(1-p_{1}\right)^{n_{1}-x} \left(1-p_{2}\right)^{n_{2}-y} t^{xy}.
\end{equation*}

\noindent We will denote (henceforth, in short) the bivariate  binomial conditionals distribution  of  the pair $\left(X,Y\right)$  with the p.m.f. in (1)  as $BBCD\left(n_{1}, n_{2},p_{1}, p_{2}, t\right).$  We list some useful results related to the normalizing  constant that will be utilized later on in deriving some structural properties.

\begin{itemize}
\item[(i)] 
\begin{eqnarray*}
K^{-1}_{B}\left(n_{1}, n_{2},p_{1}, p_{2}, 1 \right)
&=&\sum_{x=0}^{n_1}\sum_{y=0}^{n_2}\binom{n_{1}}{x} \binom{n_{2}}{y} p^{x}_{1} p^{y}_{2} \left(1-p_{1}\right)^{n_{1}-x} \left(1-p_{2}\right)^{n_{2}-y}\notag\\
&=&1.
\end{eqnarray*}

\item[(ii)] 
\begin{eqnarray*}
K^{-1}_{B}\left(n_{1}, n_{2},p_{1}, p_{2}, t \right)
&=&K^{-1}_{B}\left(n_{2}, n_{1},p_{2}, p_{1}, t \right).
\end{eqnarray*}

\item[(iii)]
\begin{eqnarray*}
K^{-1}_{B}\left(n_{1}, n_{2},p_{1}, 1, t \right) 
&=&\sum_{x=0}^{n_1}\binom{n_{1}}{x}p^{x}_{1} \left(1-p_{1}\right)^{n_{1}-x} p_2^{n_2} (1-p_2)^{n_2-n_2} t^{x n_2}\notag\\
&=&p_2^{n_2} \left(1-p_1+t^{n_2} p_1\right)^{n_1}.
\end{eqnarray*}

\item[(iv)]
\begin{equation*}
K^{-1}_{B}\left(n_{1}, n_{2},1,p_{2}, t \right) 
=p_1^{n_1} \left(1-p_2+t^{n_1} p_2\right)^{n_2}.
\end{equation*}

\item[(v)]
\begin{eqnarray*}
0<K^{-1}_{B}\left(n_{1}, n_{2},p_{1}, p_{2}, t \right)
&\leq & \min\{p^{n_{1}}_{1}\left(1- p_{2}+p_{2}t^{n_{1}}\right)^{n_{2}},  p^{n_{2}}_{2}\left(1- p_{1}+p_{1}t^{n_{2}}\right)^{n_{1}}\}\notag\\
&\leq& t^{n_{1}n_{2}}.
\end{eqnarray*}

\item[(vi)]
\begin{align*}
&\frac{d}{dp_{1}}K^{-1}_{B}\left(n_{1}, n_{2},p_{1}, 1, t \right)\notag\\
&=\sum_{x=0}^{n_1}\sum_{y=0}^{n_2}\binom{n_{1}}{x}\binom{n_{2}}{y} \Bigg( x p^{x-1}_{1}  \left(1-p_{1}\right)^{n_{1}-x}+\left(n_{1}-x\right)p^{x}_{1}  \left(1-p_{1}\right)^{n_{1}-x-1}\Bigg) p^{y}_{2}\left(1-p_{2}\right)^{n_{2}-y} t^{xy}\notag\\
&=p^{-1}_{1} \sum_{x=0}^{n_1}\sum_{y=0}^{n_2} x \binom{n_{1}}{x}\binom{n_{2}}{y}p^{x}_{1} p^{y}_{2} \left(1-p_{1}\right)^{n_{1}-x} \left(1-p_{2}\right)^{n_{2}-y} t^{xy}\notag\\
&+\left(1-p_{2}\right)^{-1}\sum_{x=0}^{n_1}\sum_{y=0}^{n_2}\left(n_{1}-x\right)\binom{n_{1}}{x}\binom{n_{2}}{y}p^{x}_{1} p^{y}_{2} \left(1-p_{1}\right)^{n_{1}-x} \left(1-p_{2}\right)^{n_{2}-y} t^{xy}\notag\\
&=p^{-1}_{1}  \frac{E(X)}{K_{B}}+\left(1-p_{1}\right)^{-1}  \frac{E\left(n-X\right)}{K_{B}}.
\end{align*}

Note that the above result immediately implies that
\begin{equation*}
\frac{d}{dp_{1}}\log K^{-1}_{B}= \frac{E(X)}{p_{1}}+\frac{E\left(n-X\right)}{1-p_{1}}.
\end{equation*}

\noindent Next, observe that for $t=1,$ this reduces to 
\begin{equation*}
0=\frac{E(X)}{p_1}+\frac{E\left(n-X\right)}{1-p_1}
\implies E\left(n-X\right)=\left(p^{-1}_{1}-1\right)E(X).
\end{equation*}

\item[(vii)] Again, by re-writing the expression for  $K_{B}\left(n_{1}, n_{2},p_{1}, p_{2}, t \right)$ as

\begin{equation*}
K^{-1}_{B}\left(n_{1}, n_{2},p_{1}, 1, t \right)
=\left(1-p_{1}\right)^{n_{1}}\left(1-p_{2}\right)^{n_{2}}\sum_{x=0}^{n_1}\sum_{y=0}^{n_2}\binom{n_{1}}{x}\left(\frac{p_{1}}{1-p_{1}}\right)^{x}\binom{n_{2}}{y}\left(\frac{p_{2}}{1-p_{2}}\right)^{y} t^{xy},
\end{equation*}

\noindent and by writing\\ $$\frac{p_{1}}{1-p_{1}}=q_{1},\frac{p_{2}}{1-p_{2}}=q_{2},\,\text{and}\, S\left( n_{1}, n_{2}, q_{1}, q_{2}, t \right)=\sum_{x=0}^{n_1}\sum_{y=0}^{n_2}\binom{n_{1}}{x}q_{1}^{x}\binom{n_{2}}{y}q_{2}^{y} t^{xy},$$ we have

\begin{equation*}
K^{-1}_{B}\left(n_{1}, n_{2},p_{1}, 1, t \right)
=\left(1-p_{1}\right)^{n_{1}}\left(1-p_{2}\right)^{n_{2}}S\left( n_{1}, n_{2},  q_{1}, q_{2},  t \right).
\end{equation*}
\end{itemize}

\noindent Note that

\begin{enumerate}
\item The marginal  p.m.f. of $X$ will be

\begin{eqnarray}
P\left(X=x\right)
&=&K_{B}\left(n_{1}, n_{2},p_{1}, 1, t \right)\binom{n_{1}}{x}p^{x}_{1}\left(1-p_{1}\right)^{n_{1}-x} \displaystyle \sum_{y=0}^{n_2}\binom{n_{2}}{y}\left( t^{x} p_{2}\right)^{y}  \left(1-p_{2}\right)^{n_{2}-y}\notag\\
&=&K_{B}\left(n_{1}, n_{2},p_{1}, 1, t \right)\binom{n_{1}}{x}p^{x}_{1}\left(1-p_{1}\right)^{n_{1}-x}  \Bigg[1-p_{2}+ t^{x} p_{2}\Bigg]^{n_{2}},
\end{eqnarray}

\noindent for $x=0,1,2,\cdots, n_{1}.$

\item Similarly, the marginal  p.m.f. of  $Y$ will be

\begin{equation}
P\left(Y=y\right)
=K_{B}\left(n_{1}, n_{2},p_{1}, 1, t \right)\binom{n_{2}}{y}p^{y}_{2}\left(1-p_{2}\right)^{n_{2}-y}  \Bigg[1-p_{1}+ t^{y} p_{1}\Bigg]^{n_{1}},
\end{equation}

\noindent for $y=0,1,2,\cdots, n_{2}.$

\item For fixed  $t\in (0,1),$

$P\left(X=x, Y=y|n_1, n_2, p_{1},p_{2},t\right)=P\left(X=y, Y=x|n_2, n_1, p_{2},p_{1},t\right).$\\
Thus, if $p_{1}=p_{2}=p,\text{say}$ then
$P\left(X=x, Y=y|n_1, n_2, p, p, t\right)=P\left(X=y, Y=x| n_2, n_1, p, p, t\right).$ Furthermore,  if $n_1=n_2=n\text{say}$ then
$P\left(X=x, Y=y|n, t\right)=P\left(X=y, Y=x|n, t\right).$ 
\end{enumerate}

\smallskip

\begin{center}
\begin{figure}
	\begin{tabular}{lll}
	\hline\\[-8pt]
	\multicolumn{3}{c}{$n_1=n_2= 10$, $p_1 = 0.5$, $p_2 = 0.5$}\\[3pt]\hline\\[-8pt]
	(i) $t = 0.95$ & (ii) $t = 0.5$ & (iii) $t = 0.05$ \\
	\includegraphics[scale = 0.35]{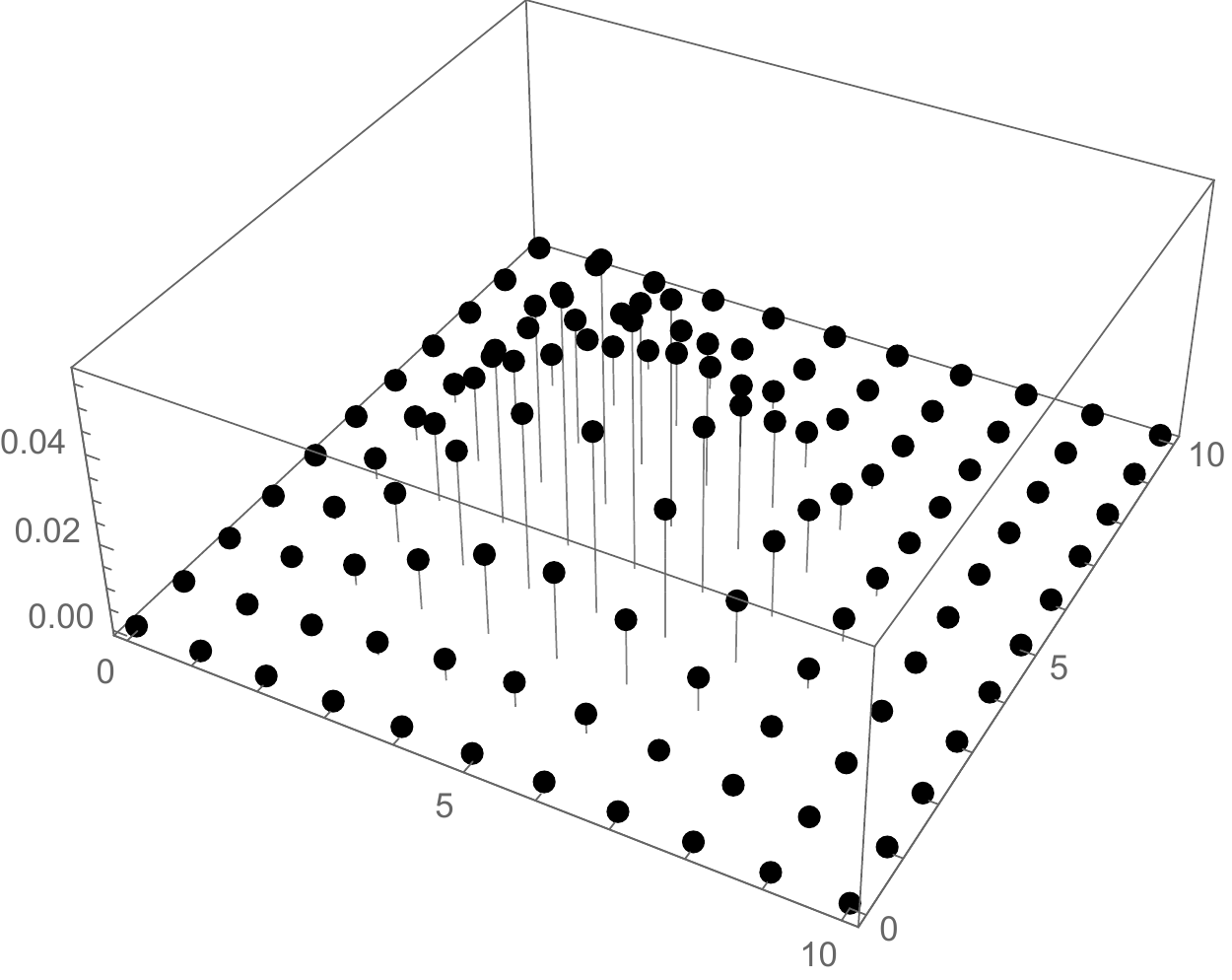} & \includegraphics[scale = 0.35]{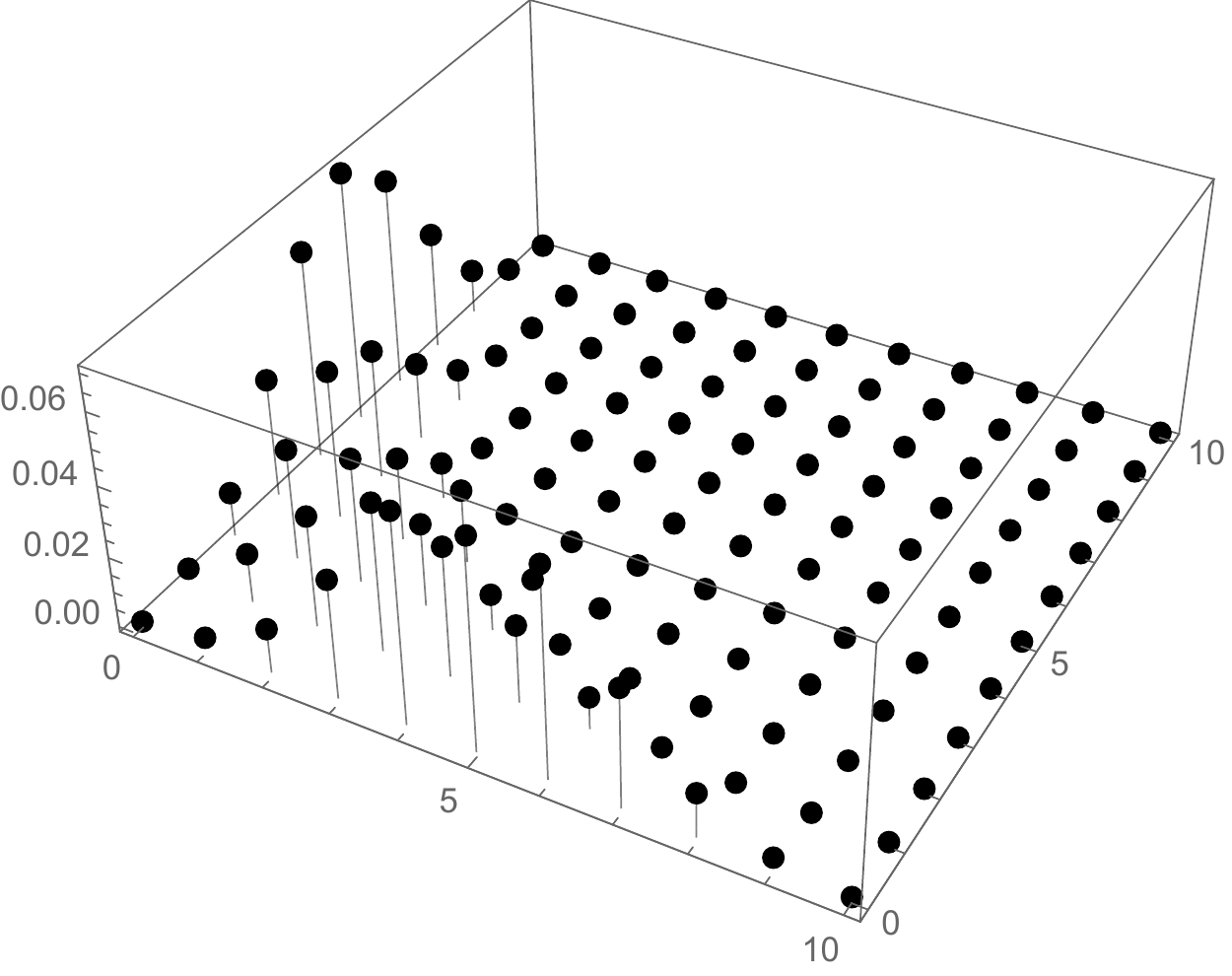}  & \includegraphics[scale = 0.35]{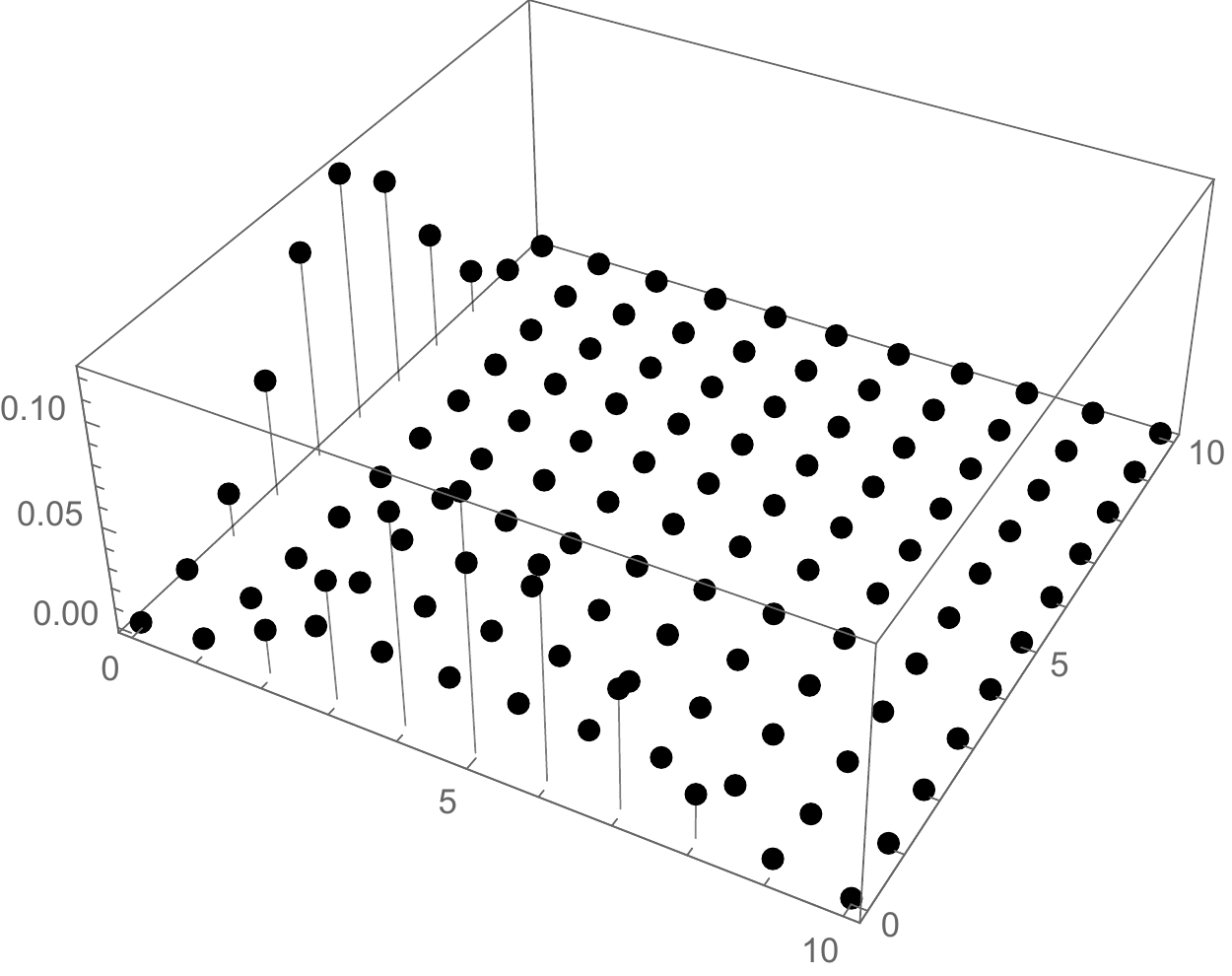} \\[3pt]\hline\\[-8pt]
	\multicolumn{3}{c}{$n_1=n_2= 20$, $p_1 = 0.1$, $p_2 = 0.1$}\\[3pt]\hline\\[-8pt]
	(iv) $t = 0.95$ & (v) $t = 0.5$ & (vi) $t = 0.05$ \\
	\includegraphics[scale = 0.35]{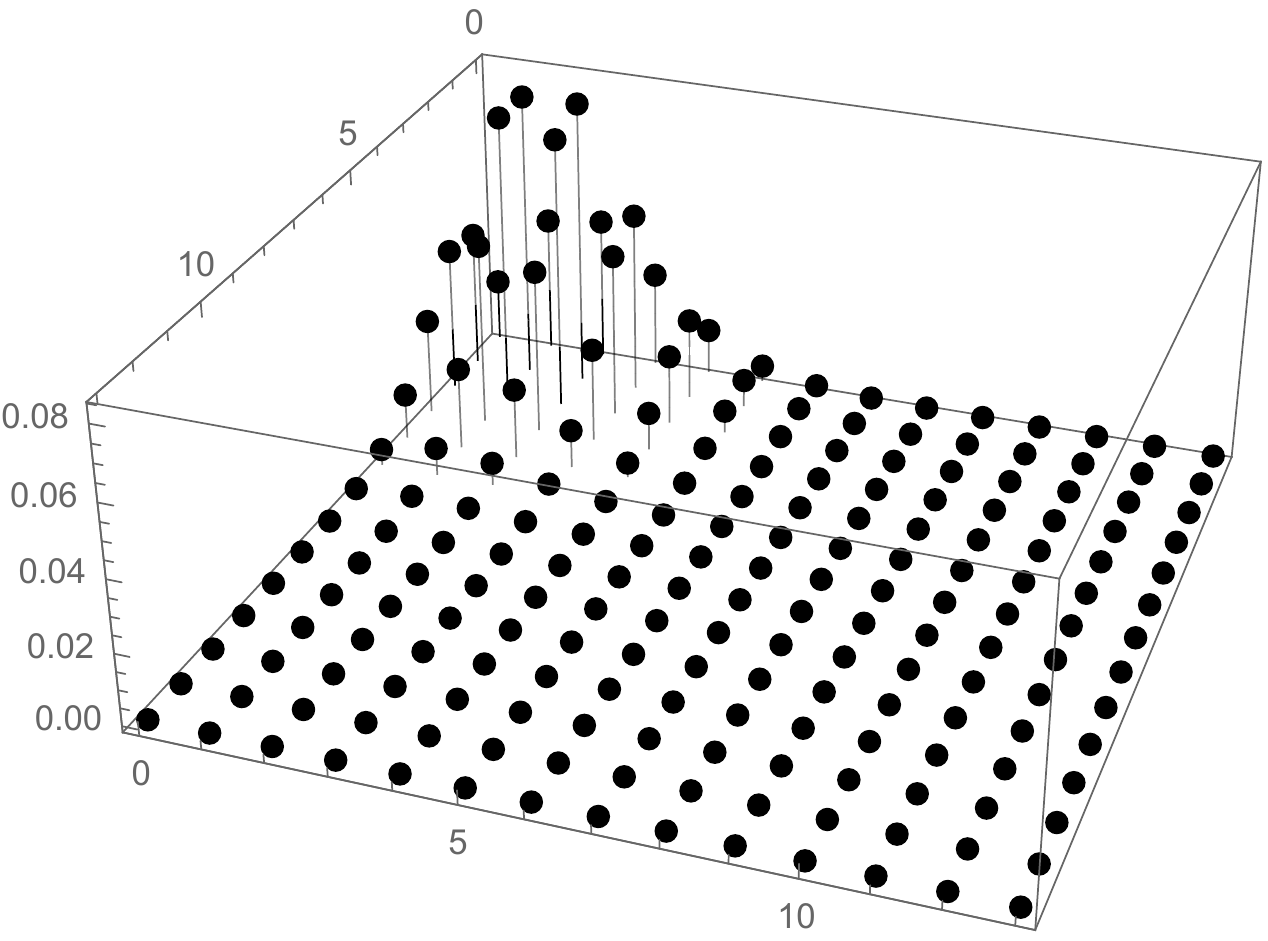} & \includegraphics[scale = 0.35]{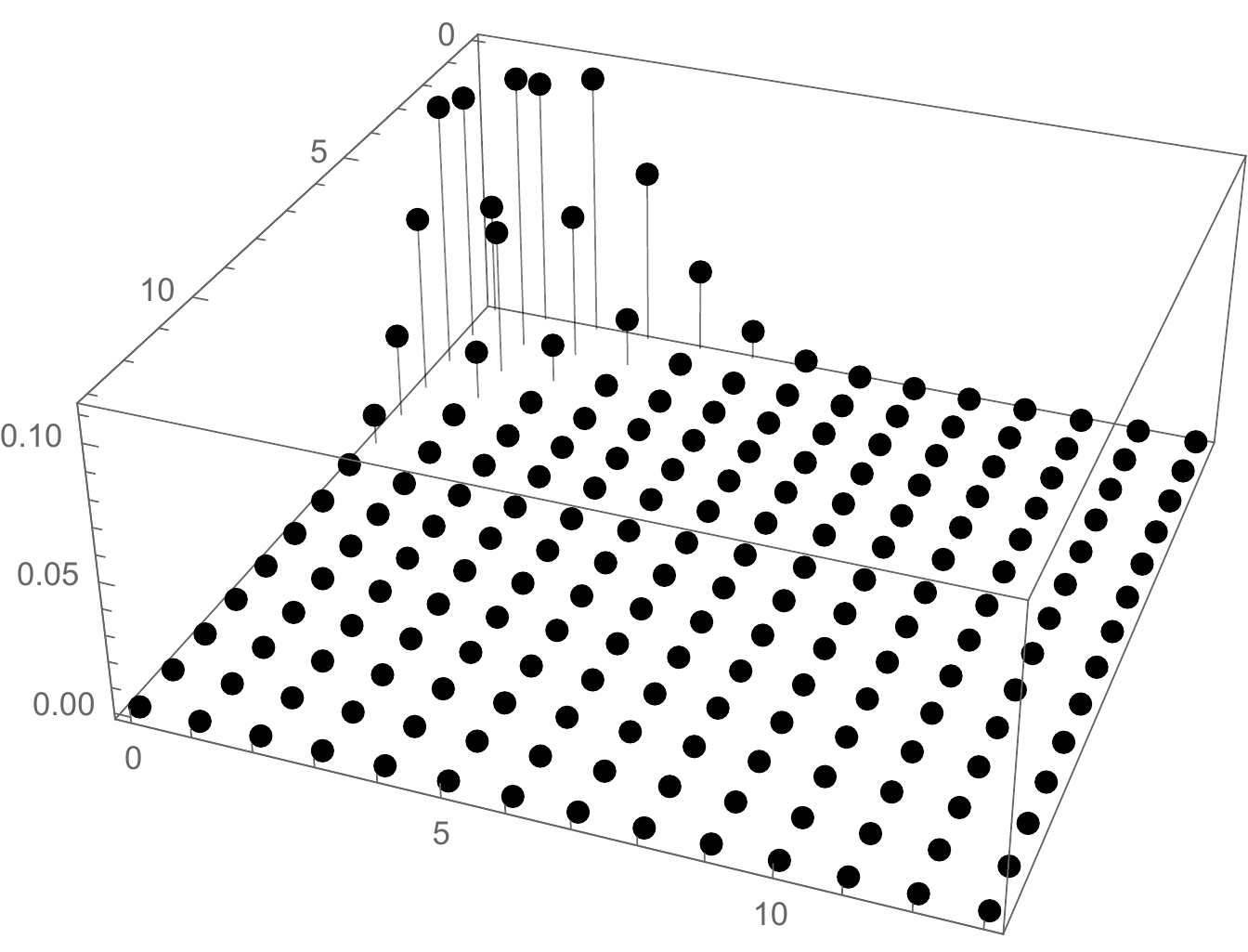}  & \includegraphics[scale = 0.35]{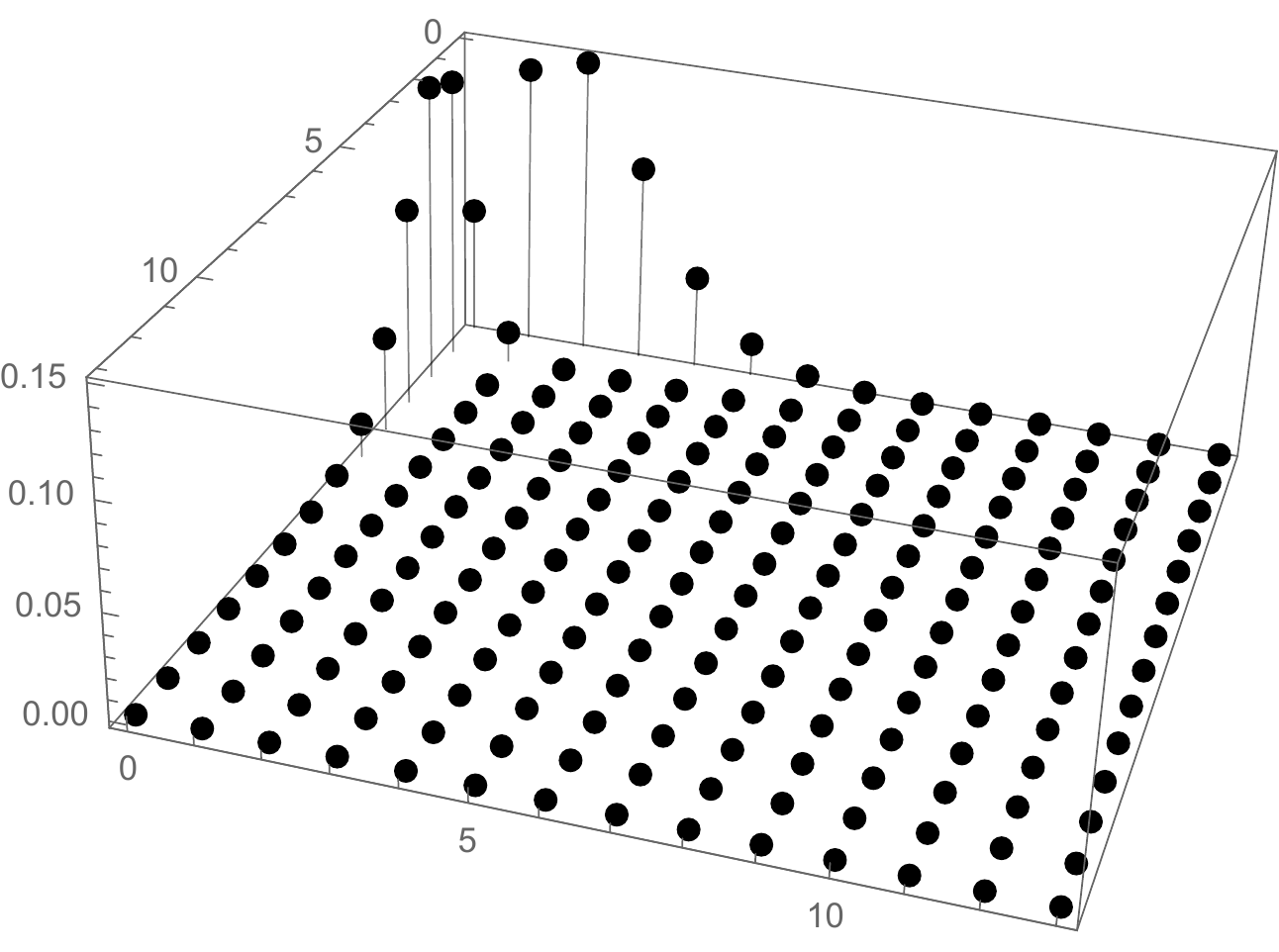} \\[3pt]\hline\\[-8pt]
	\multicolumn{3}{c}{$n_1=n_2= 30$, $p_1 = 0.9$, $p_2 = 0.9$}\\[3pt]\hline\\[-8pt]
	(vii) $t = 0.95$ & (viii) $t = 0.5$ & (ix) $t = 0.05$ \\
	\includegraphics[scale = 0.45]{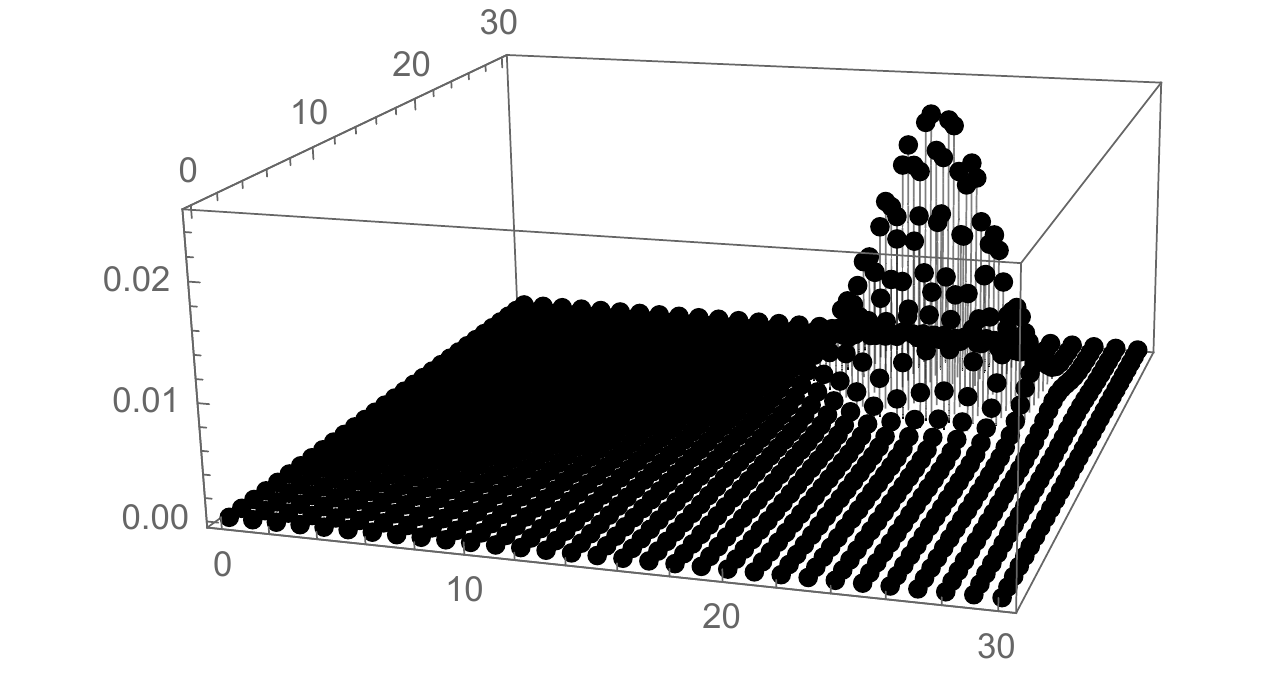} & \includegraphics[scale = 0.4]{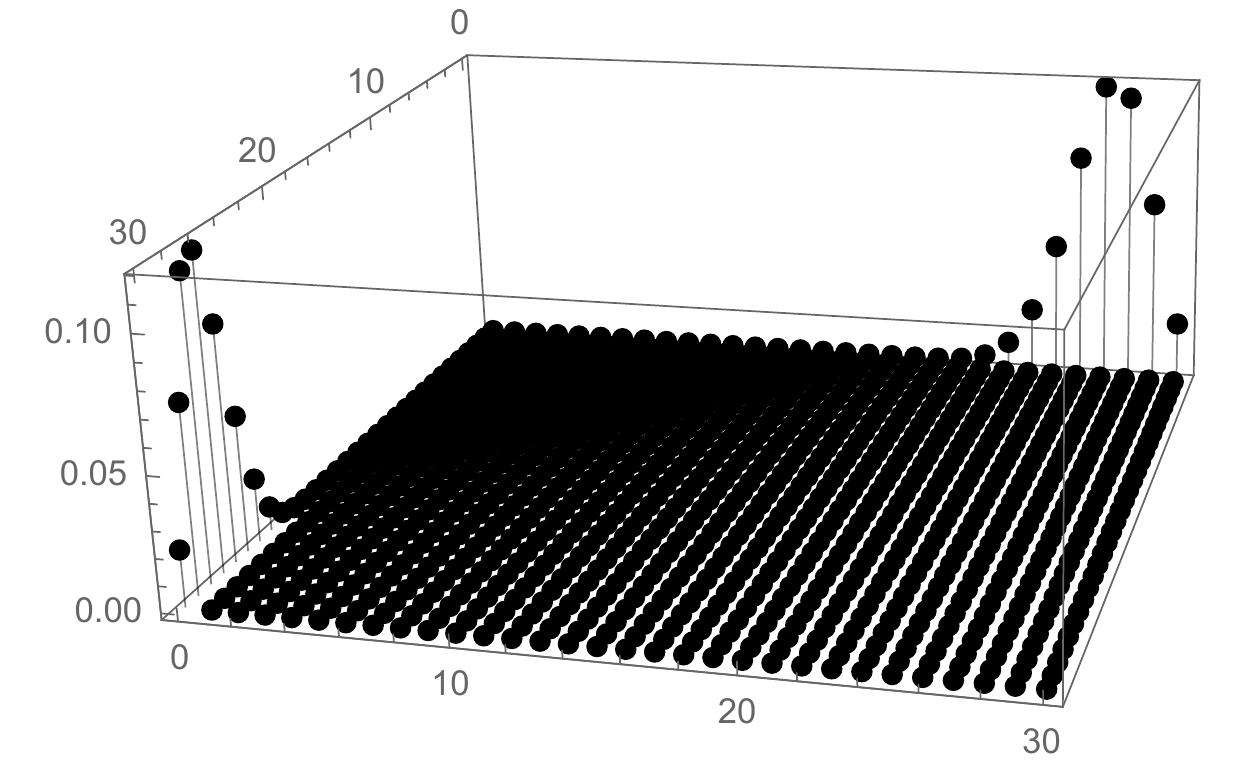}  & \includegraphics[scale = 0.35]{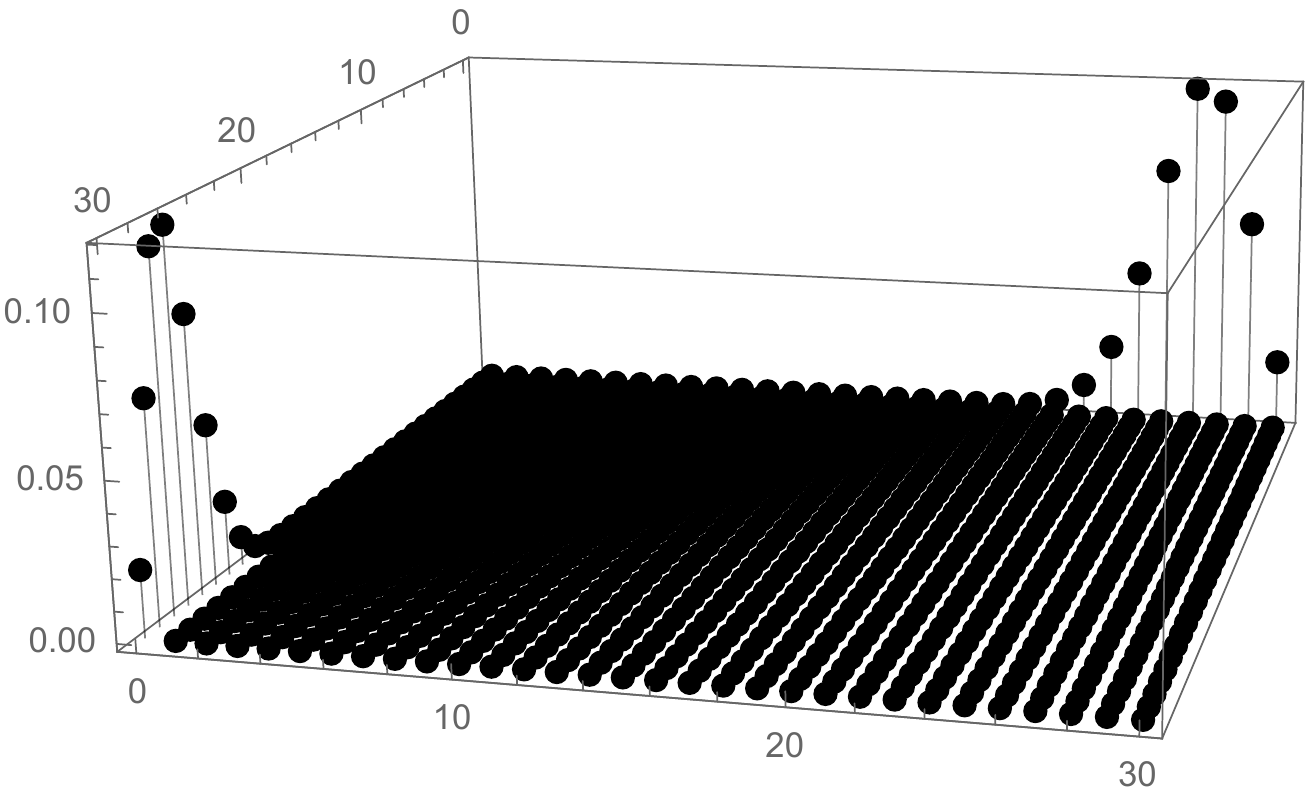} \\[3pt]\hline\\[-8pt]
	\multicolumn{3}{c}{$n_1=n_2= 15$, $p_1 = 0.3$, $p_2 = 0.7$}\\[3pt]\hline\\[-8pt]
	(x) $t = 0.95$ & (xi) $t = 0.5$ & (xii) $t = 0.05$ \\
	\includegraphics[scale = 0.4]{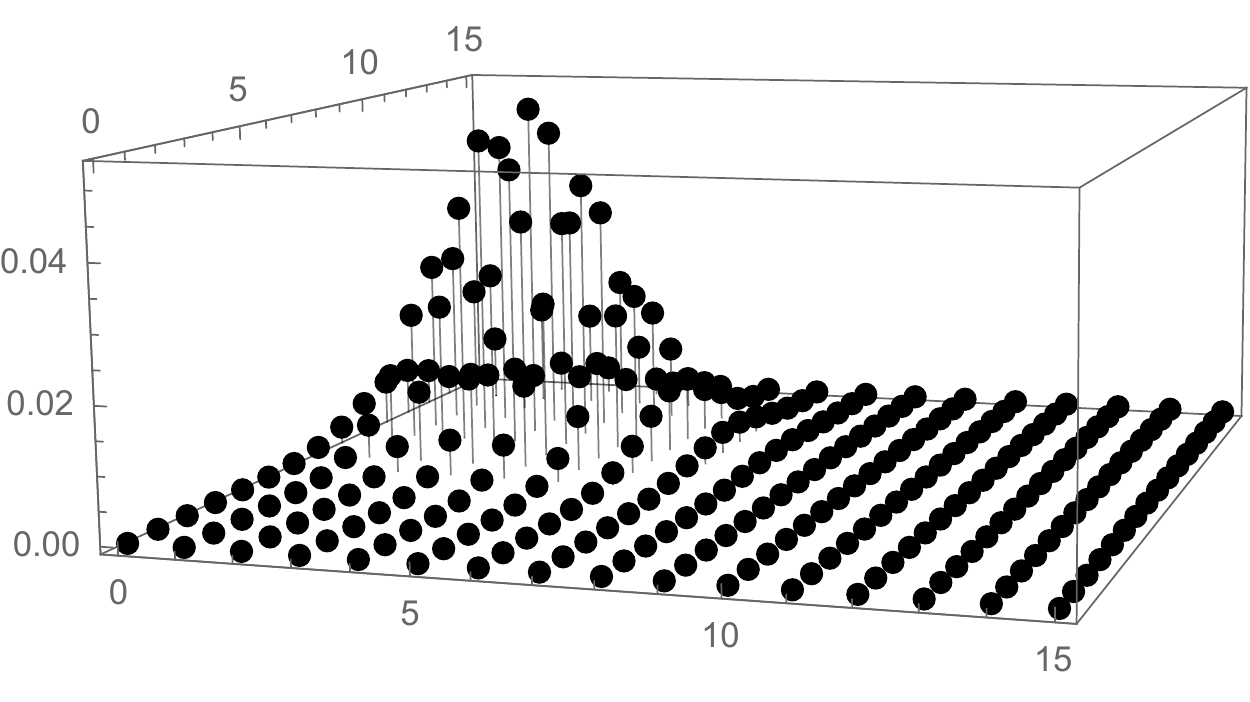} & \includegraphics[scale = 0.4]{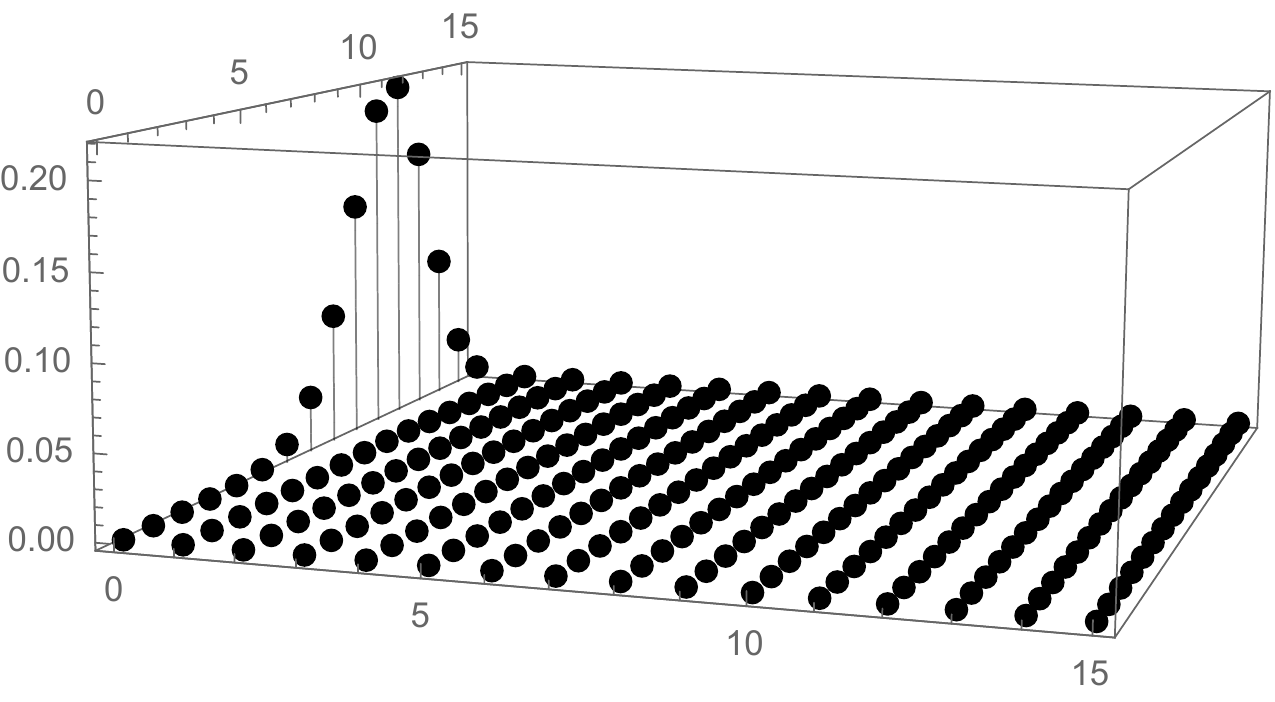}  & \includegraphics[scale = 0.4]{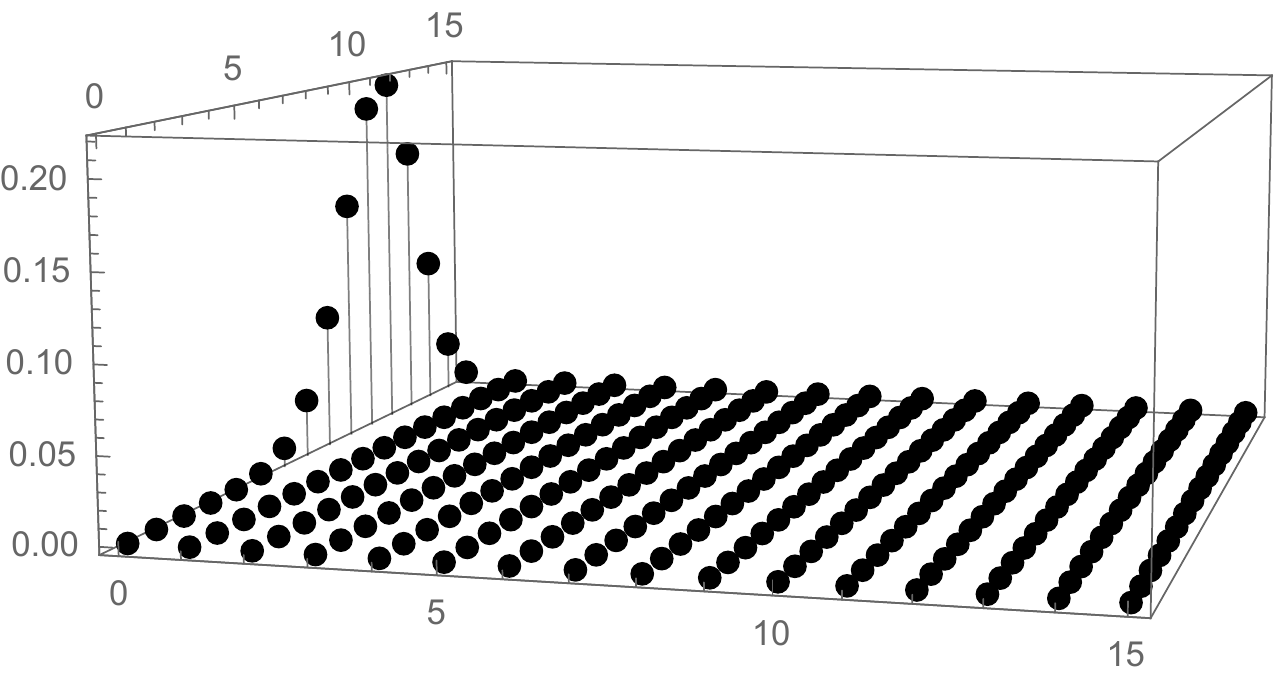} \\\hline
\end{tabular} 
\caption{Examples of p.m.f. plots for the BBCD distribution}
\label{exeplots}
\end{figure}
\end{center}

Some representative p.m.f. plots for varying parameter choices are provided in Figure \ref{exeplots}.

\section{Structural properties}
 Note that since,  $X|Y=y\sim \text{binomial}\left(n_{1}, \frac{t^{y} p_{1}}{1-p_{1}+ t^{y} p_{1}}\right),$\\ 
 $$E\left(X|Y=y\right)=n_{1}\left[\frac{t^{y} p_{1}}{1-p_{1}+ t^{y} p_{1}}\right],$$ and 
 $$Var\left(X|Y=y\right)=n_{1}\left[\frac{t^{y} p_{1}}{1-p_{1}+ t^{y} p_{1}}\right]\left[1-\frac{t^{y} p_{1}}{1-p_{1}+ t^{y} p_{1}}\right].$$
 
 \noindent Consequently,

 \begin{eqnarray}
 E\left(X\right)
 &=&E_{Y}\left(X|Y=y\right)\notag\\
 &=&K_{B}\left(n_{1}, n_{2},p_{1}, p_{2}, t \right) \sum_{y=0}^{n_2}\bigg(n_{1}\left[\frac{t^{y} p_{1}}{1-p_{1}+ t^{y} p_{1}}\right]\bigg)\binom{n_{2}}{y}p_{2}^{y}  \left(1-p_{2}\right)^{n_{2}-y}\notag\\
 &=&n_{1}K_{B}\left(n_{1}, n_{2},p_{1}, p_{2}, t \right) \sum_{y=0}^{n_2}\binom{n_{2}}{y}\left(t p_{2}\right)^{y} \left(1-p_{2}\right)^{n_{2}-y}\Bigg(\sum_{j=0}^{n_1-1}\binom{n_1-1}{j} \left(t^{y} p_{1}\right)^{j}  \left(1-p_{1}\right)^{n_{1}-1-j}\Bigg)\notag\\
 &=&n_{1}p_{1} K_{B}\left(n_{1}, n_{2},p_{1}, p_{2}, t \right)\sum_{j=0}^{n_1-1}\binom{n_1-1}{j}  p_{1}^{j}  \left(1-p_{1}\right)^{n_{1}-1-j}\Bigg(\sum_{y=0}^{n_2}\binom{n_{2}}{y}\left(p_{2}t^{j+1}\right)^{y}\left(1-p_{2}\right)^{n_{2}-y}\Bigg)\notag\\
 &=&n_{1}p_{1} K_{B}\left(n_{1}, n_{2},p_{1}, p_{2}, t \right)\sum_{j=0}^{n_1-1}\binom{n_1-1}{j}  p_{1}^{j}  \left(1-p_{1}\right)^{n_{1}-1-j}\Bigg[1-p_{2}+p_{2}t^{j+1}\Bigg]^{n_{2}}.
  \end{eqnarray}
  
  \noindent Alternatively, we can re-write (4) as
  
 \begin{eqnarray}
 E\left(X\right)
 &=& K_{B}\left(n_{1}, n_{2},p_{1}, p_{2}, t \right)\sum_{x=0}^{n_1}x \binom{n_1}{x}  p_{1}^{x}  \left(1-p_{1}\right)^{n_{1}-x}\left[1-p_{2}+p_{2}t^{x}\right]^{n_{2}}\notag\\
 &=&  K_{B}\left(n_{1}, n_{2},p_{1}, p_{2}, t \right)\sum_{x=0}^{n_1}x \binom{n_1}{x}  p_{1}^{x}  \left(1-p_{1}\right)^{n_{1}-x} w_{x},
  \end{eqnarray} 

\noindent $w_{x}=\left[1-p_{2}+p_{2}t^{x}\right]^{n_{2}}.$ Clearly,

\begin{eqnarray*}
w(x) 
&=& 1, \quad \text{for}\quad t=1,\notag\\
&<& 1, \quad \text{for}\quad t<1,\notag\\
&>& 1, \quad \text{for}\quad t>1.\notag\\
  \end{eqnarray*} 

Hence, for $0<t\leq 1,$ $E(X)\leq K_{B}\left(n_{1}, n_{2},p_{1}, p_{2}, t \right) n_{1}p_{1}.$ And for $t>1,$ $E(X)> K_{B}\left(n_{1}, n_{2},p_{1}, p_{2}, t \right) n_{1}p_{1}.$

Similarly we can show that
 \begin{eqnarray}
 E\left(Y\right)
 &=&n_{2}p_{2} K_{B}\left(n_{1}, n_{2},p_{1}, p_{2}, t \right)\sum_{j=0}^{n_{2}-1}\binom{n_{2}-1}{j}  p_{2}^{j}  \left(1-p_{2}\right)^{n_{2}-1-j}\Bigg[1-p_{1}+p_{1}t^{j+1}\Bigg]^{n_{1}}.
 \end{eqnarray}

 \medskip
 
\noindent {\bf Theorem $1.$} If $\left(X,Y\right)\sim BBCD\left(n_1, n_2, p_{1},p_{2}, t\right),$ then the correlation between $X$ and $Y$ is $< 0, >0, = 0$ respectively for $ t < 0, > 0, = 0$. \\
\noindent \textit{Proof.} \\
We consider the case when $t<0$ to show negative correlation.
\begin{align*}
&E(X)E(Y)\notag\\
&= \Bigg[n_{1}p_{1} K_{B}\left(n_{1}, n_{2},p_{1}, p_{2}, t \right)\sum_{i=0}^{n_1-1}\binom{n_1-1}{i}  p_{1}^{i}  \left(1-p_{1}\right)^{n_{1}-1-j}\Bigg[1-p_{2}+p_{2}t^{j+1}\Bigg]^{n_{2}}\Bigg]\notag\\
&\times \Bigg[n_{1}p_{1} K_{B}\left(n_{1}, n_{2},p_{1}, p_{2}, t \right)\sum_{j=0}^{n_2-1}\binom{n_2-1}{j}  p_{2}^{j}  \left(1-p_{2}\right)^{n_{2}-1-j}\Bigg[1-p_{1}+p_{1}t^{j+1}\Bigg]^{n_{1}}\Bigg]\notag\\
&=\Bigg(K_{B}\left(n_{1}, n_{2},p_{1}, p_{2}, t \right)\Bigg)^{2}\bigg(\sum_{j=0}^{n_2-1}\binom{n_2-1}{j}  p_{2}^{j}  \left(1-p_{2}\right)^{n_{2}-1-j}\Bigg[1-p_{1}+p_{1}t^{j+1}\Bigg]^{n_{1}}\Bigg]\bigg)\notag\\
&\times \Bigg[n_{1} n_{2}p_{1} p_{2} t \sum_{i=0}^{n_1-1}\binom{n_1-1}{i} (t p_{1})^{i}  \left(1-p_{1}\right)^{n_{1}-1-j}\Bigg[1-p_{2}+p_{2}t^{j+1}\Bigg]^{n_{2}-1}\left\{\frac{1-p_{2}+p_{2}t^{i+1}}{t^{i+1}}\right\}\Bigg]\notag\\
& > E\left(XY\right)\Bigg[K_{B}\left(n_{1}, n_{2},p_{1}, p_{2}, t \right)\sum_{j=0}^{n_2-1}\binom{n_2-1}{j}  p_{2}^{j}  \left(1-p_{2}\right)^{n_{2}-1-j}\left(1-p_{1}+p_{1}t^{j+1}\right)^{n_1}\Bigg]\notag\\
&\hskip6cm \text{[Since} \hskip.1cm 1-p_{2}+p_{2}t^{i+1}>t^{i+1} \hskip.1cm \text{for} \hskip.1cm t<1].\notag\\
&\text{Again for}\hskip.1cm t<1,\notag\\
&\sum_{j=0}^{n_2-1}\binom{n_2-1}{j}  p_{2}^{j}  \left(1-p_{2}\right)^{n_{2}-1-j}\left(1-p_{1}+p_{1}t^{j+1}\right)^{n_1}\notag\\
&=\sum_{i=0}^{n_1}\binom{n_1}{i}  p_{1}^{i}  \left(1-p_{1}\right)^{n_{1}-i}
\sum_{j=0}^{n_2-1}\binom{n_2-1}{j}  p_{2}^{j}  \left(1-p_{2}\right)^{n_{2}-1-j} t^{ij+i}\notag\\
&>\sum_{i=0}^{n_1}\binom{n_1}{i}  p_{1}^{i}  \left(1-p_{1}\right)^{n_{1}-i}
\sum_{j=0}^{n_2-1}\binom{n_2-1}{j}  p_{2}^{j}  \left(1-p_{2}\right)^{n_{2}-1-j} t^{ij}\,\,\text{since}\,\, t^i<1\notag\\
&=K_{B}^{-1}\left(n_{1}, n_{2}-1,p_{1}, p_{2}, t \right).\notag\\
\end{align*}
Therefore
\begin{eqnarray*}
E(X)E(Y)&>&  E\left(XY\right)\frac{K_{B}^{-1}\left(n_{1}, n_{2}-1,p_{1}, p_{2}, t \right)}{K_{B}^{-1}\left(n_{1}, n_{2},p_{1}, p_{2}, t \right)}
  \notag\\
&>&E\left(XY\right),\text{Since}\,\,
K_{B}^{-1}\left(n_{1}, n_{2},p_{1}, p_{2}, t \right)\,\, \text{is a decreasing function of}\,\, n_2. 
\end{eqnarray*}
Hence $Cov(X,Y)=E(XY)-E(X)E(Y)<0.$

\medskip{}

\noindent {\bf Theorem $2.$} If $\left(X,Y\right)\sim BBCD\left(n_1, n_2, p_{1},p_{2}, t\right),$ then the joint the joint factorial moment  will be given by
\begin{eqnarray*}
E\left(X_{\left(r\right)}Y_{\left(s\right)}\right)
&=&\frac{t^{-rs}n_{1(r)}n_{2(s)}p^{r}_{1}p^{s}_{2} \left(1-p_{1}\right)^{n_{1}-r}\left(1-p_{2}\right)^{n_{2}-s}}
{ \left(1-p_{1}\right)^{n_{1}}\left(1-p_{2}\right)^{n_{2}} S\left( n_{1}, n_{2}, \frac{p_{1}}{1-p_{1}}, \frac{p_{2}}{1-p_{2}},  t \right)}\notag\\
&&\times S\left( n_{1}-r, n_{2}-s, \frac{ t^{s} p_{1}}{1-p_{1}}, \frac{ t^{r} p_{2}}{1-p_{2}},  t \right).
\end{eqnarray*}

\noindent \textit{Proof.} Simple and thus excluded. \\

Observe that for $t=1,$ $E\left(X_{\left(r\right)}Y_{\left(s\right)}\right)=n_{1(r)}n_{2(s)} p^{r}_{1}p^{s}_{2}.$
Putting $r=1,s=1$ we get
\begin{eqnarray*}
E\left(XY\right)
&=&\frac{t^{-1}n_{1(r)}n_{2(s)}p_{1}p_{2} \left(1-p_{1}\right)^{-1}\left(1-p_{2}\right)^{-1}}
{ S\left( n_{1}, n_{2}, \frac{p_{1}}{1-p_{1}}, \frac{p_{2}}{1-p_{2}},  t \right)}\notag\\
&&\times S\left( n_{1}-1, n_{2}-1, \frac{ t p_{1}}{1-p_{1}}, \frac{ t p_{2}}{1-p_{2}},  t \right).
\end{eqnarray*}

\bigskip

\noindent {\bf Theorem $3.$} If $\left(X,Y\right)\sim BBCD\left(n_{1}, n_{2},p_{1},p_{2}, t\right),$ then the joint probability generating function (p.g.f.)  will be given  by
\begin{eqnarray}
   G_{X,Y}\left(s_1, s_2\right)
   &=&E\left(s^{X}_{1} s^{Y}_{2}\right)\notag\\
   &=&\frac{\sum_{x=0}^{n_1}\sum_{y=0}^{n_2} s^{x}_{1} s^{y}_{2} \binom{n_{1}}{x}\binom{n_{2}}{y}p^{x}_{1} p^{y}_{2} \left(1-p_{1}\right)^{n_{1}-x} \left(1-p_{2}\right)^{n_{2}-y} t^{xy}}
   {\sum_{x=0}^{n_1}\sum_{y=0}^{n_2} s^{x}_{1} s^{y}_{2} \binom{n_{1}}{x}\binom{n_{2}}{y}p^{x}_{1} p^{y}_{2} \left(1-p_{1}\right)^{n_{1}-x} \left(1-p_{2}\right)^{n_{2}-y} t^{xy}} \notag\\
   &=& \frac{\sum_{x=0}^{n_1}\sum_{y=0}^{n_2} \left(\frac{s_{1}p_{1}}{1-p_{1}}\right)^{x}\left(\frac{s_{2}p_{2}}{1-p_{2}}\right)^{y}t^{xy}}
   {\sum_{x=0}^{n_1}\sum_{y=0}^{n_2} \left(\frac{p_{1}}{1-p_{1}}\right)^{x}\left(\frac{p_{2}}{1-p_{2}}\right)^{y}t^{xy}}\notag\\
   &=&\frac{S\bigg(n_{1}, n_{2},\frac{s_{1}p_{1}}{1-p_{1}}, \frac{s_{2}p_{2}}{1-p_{2}},t\bigg)}{S\bigg(n_{1}, n_{2},\frac{p_{1}}{1-p_{1}}, \frac{p_{2}}{1-p_{2}},t\bigg)}.  
   \end{eqnarray}

\smallskip
\noindent  for $0<s_1<1, \quad 0<s_{2}<1.$ \\

\medskip

\noindent Therefore, the joint moment generating function (m.g.f.) of $\left(X,Y\right)$ will be

\begin{equation*}
  M_{X,Y}\left(t_1, t_2\right)=\frac{S\bigg(n_{1}, n_{2},\frac{\exp(t_{1})p_{1}}{1-p_{1}}, \frac{\exp(t_{1})p_{2}}{1-p_{2}},t\bigg)}{S\bigg(n_{1}, n_{2},\frac{p_{1}}{1-p_{1}}, \frac{p_{2}}{1-p_{2}},t\bigg)},
\end{equation*}

\noindent for $|t_1|<1, \quad |t_2|<1.$

\noindent In particular, for $t=1,$ $S\left(n_{1}, n_{2},q_{1}, q_{2}, 1\right)=\left(1+q_{1}\right)^{n_{1}}\left(1+q_{2}\right)^{n_{2}}.$ Then, (7) reduces to 

\begin{eqnarray*}
   G_{X,Y}\left(s_1, s_2\right)
&=&\frac{\left(1+\frac{s_{1}p_{1}}{1-p_{1}}\right)^{n_1} \left(1+\frac{s_{2}p_{2}}{1-p_{2}}\right)^{n_2} }{\left(1+\frac{p_{1}}{1-p_{1}}\right)^{n_1} \left(1+\frac{p_{2}}{1-p_{2}}\right)^{n_2}}\notag\\
&=&\left(1+s_{1}p_{1}-p_{1}\right)^{n_1} \left(1+s_{2}p_{2}-p_{2}\right)^{n_2}.
   \end{eqnarray*}

\bigskip

\noindent {\bf Theorem $4.$}  If $\left(X,Y\right)\sim BBCD\left(n_{1}, n_{2},p_{1},p_{2}, t\right),$ then

\begin{eqnarray*}
P\left(X=m, Y=n\right)
&=&t^{n} p_{1}\left(1-p_{1}\right)^{-1} \Bigg[\frac{n_1-m+1}{m}\Bigg]\times P\left(X=m-1, Y=n\right), \quad \text{for}\quad m\geq 1\notag\\
&=&t^{m} p_{2}\left(1-p_{2}\right)^{-1} \Bigg[\frac{n_2-n+1}{n}\Bigg]\times P\left(X=m, Y=n-1\right), \quad \text{for}\quad n\geq 1\notag\\
&=& p_{1}p_{2} P\left(X=m-1, Y=n-1\right)\Bigg[\frac{n_1-m+1}{m}\Bigg]\times \Bigg[\frac{n_2-n+1}{n}\Bigg], \quad \text{for}\quad (m,n)\geq 1.
\end{eqnarray*}

\medskip

 \noindent \textit{Proof.} Simple and thus excluded.

 \medskip
 
 \noindent {\bf Theorem $5.$} $\left(X,Y\right)\sim BBCD\left(n_{1}, n_{2},p_{1},p_{2}, t\right),$ belongs to three parameter exponential family.

\bigskip

\textit{Proof.}
The joint p.m.f. of $\left(X,Y\right)\sim BBCD\left(n_{1}, n_{2},p_{1},p_{2}, t\right),$ will be a member of the $3$-parameter exponential family  if its p.m.f. can be expressed in the form

\begin{equation}
h\left(x,y\right)=\exp \left[\sum_{j=1}^{3} {\delta_{j}\left(p_{1}, p_{2}, t\right)W_j(x,y)} - M\left(p_{1},p_{2}, t\right)\right].
\end{equation}

\noindent From  (1), it is easy to observe that the proposed distribution belongs to the exponential family by rewriting the p.m.f. as

\begin{equation}
P\left(X=x, Y=y\right)
=\exp \exp \Bigg[\sum_{j=1}^{3} {\delta_{j}\left(p_{1}, p_{2}, t\right)W_j(x,y)} - M\left(p_{1},p_{2}, t\right)\Bigg],
\end{equation}
and identifying

\begin{eqnarray*}
&&M\left(p_{1},p_{2}, t\right)=\log K_{B}\left(n_{1}, n_{2},p_{1}, p_{2}, t \right)+n_{1} \log \left(1-p_{1}\right)+n_{2} \log \left(1-p_{2}\right)+\log \left( \binom{n_1}{x} \binom{n_2}{y}\right);\\
&&W_{1}(x,y)=x; W_{2}(x,y)=y; W_{3}(x,y)=xy;\\
&& \delta_{1}\left(p_{1},p_{2},t\right)=\log \frac{p_{1}}{1-p_{1}}; \delta_{2}\left(p_{1},p_{2},t\right)=\log \frac{p_{2}}{1-p_{2}}; \delta_{3}\left(p_{1},p_{2},t\right)=\log t.\\
\end{eqnarray*}
Thus, based on a sample of size $m$ from BBCD, $\left(\sum x,\sum y, \sum x y\right)$ is complete sufficient for $\left(p_{1},p_{2}, t\right).$\\
Distributions belonging to exponential family enjoys many properties. For example mean, variance, co-variance and moment generating functions can be easily derived using differentiation's of $M\left(p_{1},p_{2},t\right)$. Moreover using Lehmann-Scheffe (Lehman and Scheffe (1982)) result, it may be possible to derive UMVUE of the parameters, provided we get hold of function of $T$ that is unbiased for the parameter. Even otherwise one can derive MVUE implementing bias correction.

\medskip

\noindent {\bf Theorem $6.$}  If $\left(X,Y\right)\sim BBCD\left(n_{1}, n_{2},p_{1},p_{2}, t\right),$ then

\begin{eqnarray*}
P\left(X<Y\right)
&=&\left[ K_{B}\left(n_{1}, n_{2},p_{1}, p_{2}, t \right)\right]^{2}\Bigg(\sum_{j=0}^{\min\{n_1, n_2-1\}}\sum_{x=0}^{j}\left\{\binom{n_{2}}{j+1} p^{j+1}_{2}\left(1-p_{2}\right)^{n_2-j-1}\right\}
\left\{1-p_{1}+p_{1}t^{j+1}\right\}^{n_1}\notag\\
&&\times \binom{n_1}{x}p^{x}_{1}\left(1-p_{1}\right)^{n_1-x}\left[1-p_{2}+p_{2}t^{x}\right]^{n_2}
\Bigg).
\end{eqnarray*}

\medskip

\noindent {\bf Proof.}
Observe that

\begin{align*}
&P\left(X<Y\right)\notag\\
&=\sum_{j=0}^{\min\{n_1, n_2-1\}}P\left(Y=j+1\right)P\left(X\leq j\right)\notag\\
&=\left[ K_{B}\left(n_{1}, n_{2},p_{1}, p_{2}, t \right)\right]^{2}\Bigg(\sum_{j=0}^{\min\{n_1, n_2-1\}}\sum_{x=0}^{j}\left\{\binom{n_{2}}{j+1} p^{j+1}_{2}\left(1-p_{2}\right)^{n_2-j-1}\right\}
\left\{1-p_{1}+p_{1}t^{j+1}\right\}^{n_1}\notag\\
&\times \binom{n_1}{x}p^{x}_{1}\left(1-p_{1}\right)^{n_1-x}\left[1-p_{2}+p_{2}t^{x}\right]^{n_2}\Bigg).
\end{align*}

Hence, the proof.

\medskip

{\bf Note:} Since  $0<t\leq 1,$ we may obtain a upper bound inequality of $P\left(X<Y\right)$ by setting $t=1,$ which will be as follows:

\begin{align*}
&P\left(X<Y\right)\notag\\
&\leq  \left[ K_{B}\left(n_{1}, n_{2},p_{1}, p_{2}, t \right)\right]^{2}\Bigg(\sum_{j=0}^{\min\{n_1, n_2-1\}}\sum_{x=0}^{j}\left\{\binom{n_{2}}{j+1} p^{j+1}_{2}\left(1-p_{2}\right)^{n_2-j-1}\right\}
\left\{1-p_{1}+p_{1}\right\}^{n_1}\notag\\
&\times \binom{n_1}{x}p^{x}_{1}\left(1-p_{1}\right)^{n_1-x}\Bigg).
\end{align*}

 Note that  $R=P\left(X<Y\right)$ is known as the stress- strength reliability in engineering where the random variables  $X$ and $Y$ respectively represent the stress and strength associated with a system. This measure is also useful in a probabilistic assessment of inequality in two phenomena $X$ and $Y.$ While in most cases $X$ and $Y$ are assumed to be independent in real life there may be dependence between $X$ and $Y.$ Stress-strength reliability with both variables having independent binomial distribution  was discussed in Becker et al. (2002). As such the result of Theorem $6$ has the potential to be used in such contexts.

\noindent As can be clearly seen that even though there is no closed form for the above expression of the covariance, this can be computed
by taking a large number of terms in the series above. For that the command \textit{NSum} of
Mathematica package can be used in order to get a value close to the exact.

\medskip

\noindent {\bf Theorem $7.$} If $\left(X,Y\right)\sim BBCD\left(n_{1}, n_{2},p_{1},p_{2}, t\right)$ then we have the following results.

\begin{itemize}
\item [(a)]
\begin{equation*}
P\left(X=x|X+Y=u\right)=\frac{\binom{n_{1}}{x}\binom{n_{2}}{u-x}\left[\frac{p_{1}\left(1-p_{2}\right)}{p_{2}\left(1-p_{1}\right)}\right]^{x}t^{-x^{2}}}{\sum_{x=0}^{u}\binom{n_{1}}{x}\binom{n_{2}}{u-x}\left[\frac{p_{1}\left(1-p_{2}\right)}{p_{2}\left(1-p_{1}\right)}\right]^{x}t^{-x^{2}}},
\end{equation*}

\noindent $u=0,1,\cdots, n_1+n_2.$ Here, $\max\{0, u-n_2\}\leq x\min\{u, n_{1}\},$ we write as

\begin{equation*}
P\left(X=x|X+Y=u\right)\propto \binom{n_{1}}{x}\binom{n_{2}}{u-x}\left[\frac{p_{1}\left(1-p_{2}\right)}{p_{2}\left(1-p_{1}\right)}\right]^{x}t^{-x^{2}}\binom{n_1+n_2}{u}
\left[\frac{p_{1}\left(1-p_{2}\right)}{ t^{x} p_{2}\left(1-p_{1}\right)}\right]^{x}.
\end{equation*}

Observe that for $t=1,$ we get

\begin{equation*}
P\left(X=x|X+Y=u\right)\propto \binom{n_{1}}{x}\binom{n_{2}}{u-x}\left[\frac{p_{1}\left(1-p_{2}\right)}{p_{2}\left(1-p_{1}\right)}\right]^{x}\binom{n_1+n_2}{u}
\left[\frac{p_{1}\left(1-p_{2}\right)}{ p_{2}\left(1-p_{1}\right)}\right]^{x},
\end{equation*}

\noindent which is the p.m.f. of extended hypergeometric distribution of \cite{hw}. Again, it reduces to a classical hypergeometric distribution when $p_{1}=p_{2}.$

\item [(b)] The regression of $X$ on $Y$ is given by $E\left(X|Y=y\right)=n_{1}\left\{\frac{p_{1}t^{y}}{1-p_{1}+p_{1}t^{y}}\right\},$ and 
the regression of $Y$ on $X$ is given by $E\left(Y|X=x\right)=n_{2}\left\{\frac{p_{2}t^{x}}{1-p_{2}+p_{2}t^{x}}\right\}.$
\end{itemize}

\bigskip

 \noindent \textit{Proof.} Part (a) is straight forward. Proof of part(b) can be obtained immediately by the information that states that both the conditionals are binomial with respective parameters. Precisely,
    
 \begin{itemize}
 \item  Since, $X|Y=y\sim binomial\left(q_{1}q^{y}_{3}\right),$ therefore, the regression of $X$ on $Y$ will be obtained as $X=E\left(X|Y=y\right)=n_{1}\left\{\frac{p_{1}t^{y}}{1-p_{1}+p_{1}t^{y}}\right\}.$
 \item Similarly,  since, $Y|X=x\sim binomial\left(q_{2}q^{x}_{3}\right),$ therefore, the regression of $Y$ on $X$ will be obtained as $Y=E\left(Y|X=x\right)=n_{2}\left\{\frac{p_{2}t^{x}}{1-p_{2}+p_{2}t^{x}}\right\}.$
\end{itemize}

 \bigskip

 \noindent {\bf Theorem $8.$}  If $\left(X,Y\right)\sim BBCD\left(n_{1}, n_{2},p_{1},p_{2}, t\right),$ then we have the following stochastic ordering results related to  the bivariate binomial conditional distribution in (1).

  \begin{itemize}
   \item[(a)] If $p_{1}>p_{2},$ then for any $0<t<1,$ and for fixed $n_1$ and $n_2,$ $X$
  is stochastically larger than $Y.$  This also implies that under this parametric restriction, $Y$ is smaller than $X$ in the hazard rate order, mean residual life order, and the likelihood ratio order.
  \item [(b)] If $p_{1}<p_{2},$ and for fixed $n_1$ and $n_2,$  then for any $0<t<1,$ $Y$
  is stochastically larger than $X.$ This also implies that under this parametric restriction, $X$ is smaller than $Y$ in the hazard rate order, mean residual life order, and the likelihood ratio order.

  \item[(c)] If $p_{1}<p_{2},$ and for fixed $n_1$ and $n_2,$  then for any $0<t<\min\{p_{1},p_{2}\}<1,$ $Y$
  is stochastically larger than $X.$  This also implies that under this parametric restriction, $X$ is smaller than $Y$ in the hazard rate order, mean residual life order, and the likelihood ratio order.

  \item[(d)]  If $p_{1}>p_{2},$ and for fixed $n_1$ and $n_2,$  then for any $1>t>\max\{p_{1},p_{2}\},$ $X$
  is stochastically larger than $Y.$  This also implies that under this parametric restriction, $Y$ is smaller than $X$ in the hazard rate order, mean residual life order, and the likelihood ratio order.

  \end{itemize}

\bigskip

 \noindent \textit{Proof.} The proof is quite simple and hence, the details avoided.

\medskip

\noindent {\bf Theorem $9.$} Limiting distribution:
If $\left(X,Y\right)\sim BBCD\left(n_{1}, n_{2},p_{1},p_{2}, t\right),$ then for $0<t<1,$ 

\begin{equation*}
\lim_{n_1\rightarrow\infty}\lim_{n_2\rightarrow\infty} BBCD\left(n_{1}, n_{2},p_{1},p_{2}, t\right) \overset{D}{\sim} BPD\left(\lambda_1, \lambda_2, t\right),
\end{equation*}

\noindent where as $n_1\rightarrow\infty,$ and $n_2\rightarrow\infty,$ and $p_{1}, p_{2}$ are  small such that $n_{1}p_{1}=\lambda_1, $ and $n_{2}p_{2}=\lambda_2$ are both finite positive. For details on the BPD distribution, see Ghosh et al. (2020).

\medskip

 \noindent \textit{Proof.} The proof is straightforward and hence, the details avoided.

\medskip

\noindent {\bf Theorem 10.}  Suppose  $\left(X,Y\right)\sim BBCD\left(q_{1},q_{2},q_{3}\right).$ Let $M=\max\{X,Y\},$ and $U=\min\{X,Y\}.$ Then the p.m.f. of $M$ will be

\begin{equation*}
 g_M(m)=\left(1-p_{1}\right)^{n_1} K_{B}\left(n_{1}, n_{2},p_{1}, p_{2}, t \right)  \binom{n_1}{m}\left(\frac{p_1}{1-p_1}\right)^{m} \left[1-p_{2}+t^{m}p_{2}\right]^{n_2}    
\end{equation*}

\medskip

\noindent {\bf Proof.}
\begin{eqnarray*}
P(M \leq m) &=& P\left(X \leq m, Y \leq m\right)\notag\\
&=&  \sum_{i=0}^{m}\sum_{j=0}^{m} P\left(X=i, Y=j\right)\notag\\
&=&\left(1-p_{1}\right)^{n_1} \left(1-p_{2}\right)^{n_2} K_{B}\left(n_{1}, n_{2},p_{1}, p_{2}, t \right) \sum_{i=0}^{m}\sum_{j=0}^{m} \binom{n_1}{i}\binom{n_2}{j}\left(\frac{p_1}{1-p_1}\right)^{i} \left(\frac{p_2}{1-p_2}\right)^{j}t^{ij}\notag\\
&=& \left(1-p_{1}\right)^{n_1} K_{B}\left(n_{1}, n_{2},p_{1}, p_{2}, t \right) \sum_{i=0}^{m} \binom{n_1}{i}\left(\frac{p_1}{1-p_1}\right)^{i} \left[1-p_{2}+t^{i}p_{2}\right]^{n_2}
\end{eqnarray*}
the p.m.f. of $M$ can be easily seen as 
\begin{equation*}
 g_M(m)=\left(1-p_{1}\right)^{n_1} K_{B}\left(n_{1}, n_{2},p_{1}, p_{2}, t \right)  \binom{n_1}{m}\left(\frac{p_1}{1-p_1}\right)^{m} \left[1-p_{2}+t^{m}p_{2}\right]^{n_2}    
\end{equation*}
In fact we can find the following in general result of survival function which is as follows:
\begin{eqnarray*}
 P\left(U>u\right)
 &=&P\left(X>u, Y>u\right)\notag\\
&=&\sum_{i=0}^{n_1-1-u}\sum_{j=0}^{n_2-1-u}P\left(X=u+i+1, Y=u+j+1\right)\notag\\
&=&K_{B}\left(n_{1}, n_{2},p_{1}, p_{2}, t \right)\sum_{i=0}^{n_1-1-u}\sum_{j=0}^{n_2-1-u}\binom{n_1}{u+i+1}\binom{n_2}{u+j+1}p^{u+i+1}_{1}\\
&&p^{u+i+1}_{2}\left(1-p_{1}\right)^{n_1-u-i-1}\left(1-p_{2}\right)^{n_1-u-i-1}t^{(u+i+1)^{2}}.
\end{eqnarray*}

\section{Statistical Inference}
\subsection{From sample proportions}
From the pmf of BBCD in (\ref{pmf}) we can write
\begin{eqnarray}
f_{0,0}&=&K_{B}\left(n_{1}, n_{2},p_{1}, p_{2}, t \right)\left(1-p_{1}\right)^{n_{1}} \left(1-p_{2}\right)^{n_{2}},\notag\\
f_{0,1}&=&K_{B}\left(n_{1}, n_{2},p_{1}, p_{2}, t \right) n_{2} \left(1-p_{1}\right)^{n_{1}} \left(1-p_{2}\right)^{n_{2}-1},\notag\\
f_{1,0}&=&K_{B}\left(n_{1}, n_{2},p_{1}, p_{2}, t \right) n_{1} \left(1-p_{1}\right)^{n_{1}-1} \left(1-p_{2}\right)^{n_{2}},\notag\\
f_{1,1}&=&K_{B}\left(n_{1}, n_{2},p_{1}, p_{2}, t \right) n_{1} n_{2} \left(1-p_{1}\right)^{n_{1}-1} \left(1-p_{2}\right)^{n_{2}-1}t.
\end{eqnarray}
From the above three equations we get
\begin{eqnarray}
\dfrac{f_{0,0}}{f_{0,1}}&=&\dfrac{1-p_2}{n_2 p_2},\notag\\
\dfrac{f_{0,0}}{f_{1,0}}&=&\dfrac{1-p_1}{n_1 p_1},\notag\\
\dfrac{f_{1,1}}{f_{0,1}}&=&\dfrac{n_1 t}{1-p_1}.\notag\\
\end{eqnarray}
Solving  for parameters we get
$$\widehat{p_2}=\dfrac{f_{0,1}}{f_{0,1}+n_2 f_{0,0}},$$
$$\widehat{p_1}=\dfrac{f_{1,0}}{f_{1,0}+n_2 f_{0,0}},$$ 
$$\widehat{t}=\dfrac{1-\widehat{p_1}}{n_1}\dfrac{f_{1,1}}{f_{0,1}}.$$

\subsection{Maximum Likelihood Estimation}
\label{sec41}

In this subsection, we consider the maximum likelihood estimation of the unknown parameters  $n_{1}$, $n_{2}$, $p_{1}$, $p_{2}$ and $t$ of the BBCD distribution based on a observed  sample of size $m$, $\mathbb{C}=\left((x_1, y_1), \cdots, (x_m, y_m)\right)$. For $\Delta=\left(n_1,n_2,p_{1},p_{2},t\right)$, the log-likelihood function is given by
\begin{eqnarray*}
	\ell\left(\Delta|\mathbb{C}\right)
	&=&m K_{G}\left(q_{1},q_{2},q_{3}\right)+\sum_{i=1}^m\left(\log\left[\binom{n_1}{x_i}\right]+\log\left[\binom{n_2}{y_i}\right]\right) + \log (p_{1})\sum_{i=1}^{m}x_{i}
	+\log (p_{2})\sum_{i=1}^{m}y_{i}\\
	& &  \log (1-p_{1})\sum_{i=1}^{m}(n-x_{i}) +\log (1-p_{2})\sum_{i=1}^{m}(n-y_{i}) 
	+\log (t)\sum_{i=1}^{m}x_{i}y_{i}.
\end{eqnarray*}
The maximum likelihood estimators of the unknown parameters can be obtained by maximization of the log-likelihood function, with respect to $\Delta$, but it is quite difficult or even impossible, in this case, to obtained explicit forms for the estimators, even when the values of $n_1$ and $n_2$ are assumed to be known. To overcome this problem there are several numerical methods that may be used, for example, based on the Newton-Raphson method or on the expectation–maximization (EM) algorithm. 
In this work, we have decided to used \texttt{Mathematica} software and the \texttt{NMaximize} function  to obtain the parameters estimates. This allowed us to make the estimation of the parameters $p_1$, $p_2$, $t$ assuming the values of $n_1$ and $n_2$ known, or even to make the estimation of all the parameters using as assumptions: $n_1\in \mathbb{N}$, $n_2\in \mathbb{N}$ and $p_1$, $p_2$, $t$ $\in (0,1)$.

\section{Simulation}

\begin{center}
	\begin{figure}[h]
		\includegraphics[scale = 1]{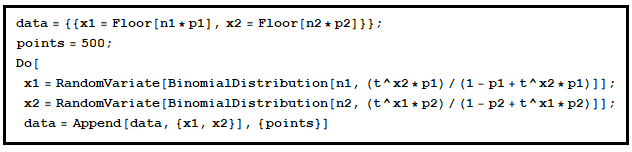} 
		\caption{Gibbs sampler method for the BBCD distribution}
		\label{gibbs}
	\end{figure}
\end{center}

Since it is not easy to simulate form the BBCD distribution we consider, in this section, the  Gibbs sampler method, see Gelfand (2000).  Using the conditional distributions

$$X|Y=y\sim \text{binomial}\left(n_{1}, \frac{t^{y} p_{1}}{1-p_{1}+ t^{y} p_{1}}\right)~~{\rm and}~~ Y|X=x\sim \text{binomial}\left(n_{2}, \frac{t^{x} p_{2}}{1-p_{2}+ t^{x} p_{2}}\right),$$
the implementation of this method is quite straightforward and the code, for the Mathematica software, to simulated $500$ points is provided in Figure \ref{gibbs}.

\begin{center}
	\begin{figure}[h]
		\begin{tabular}{ll}
			(i) & (ii)\\
		\includegraphics[scale = 0.45]{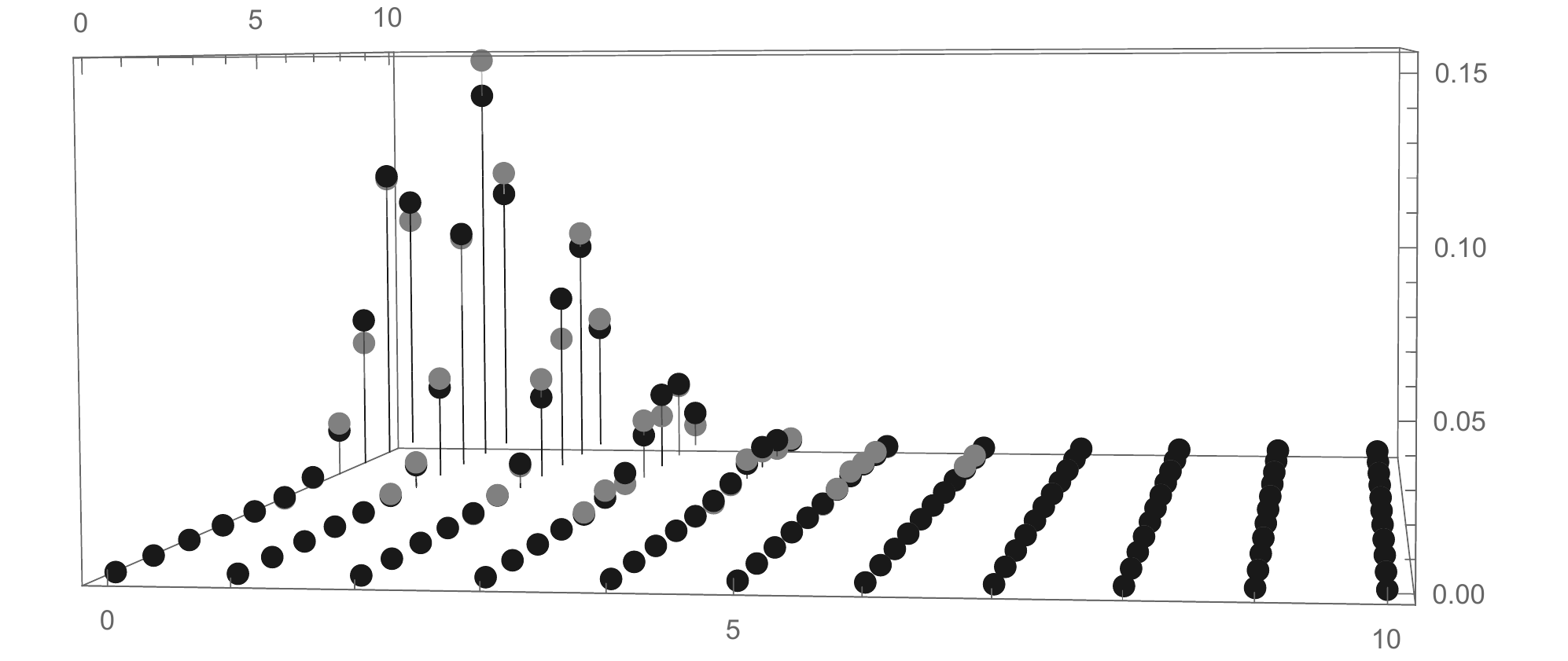} & \includegraphics[scale = 0.45]{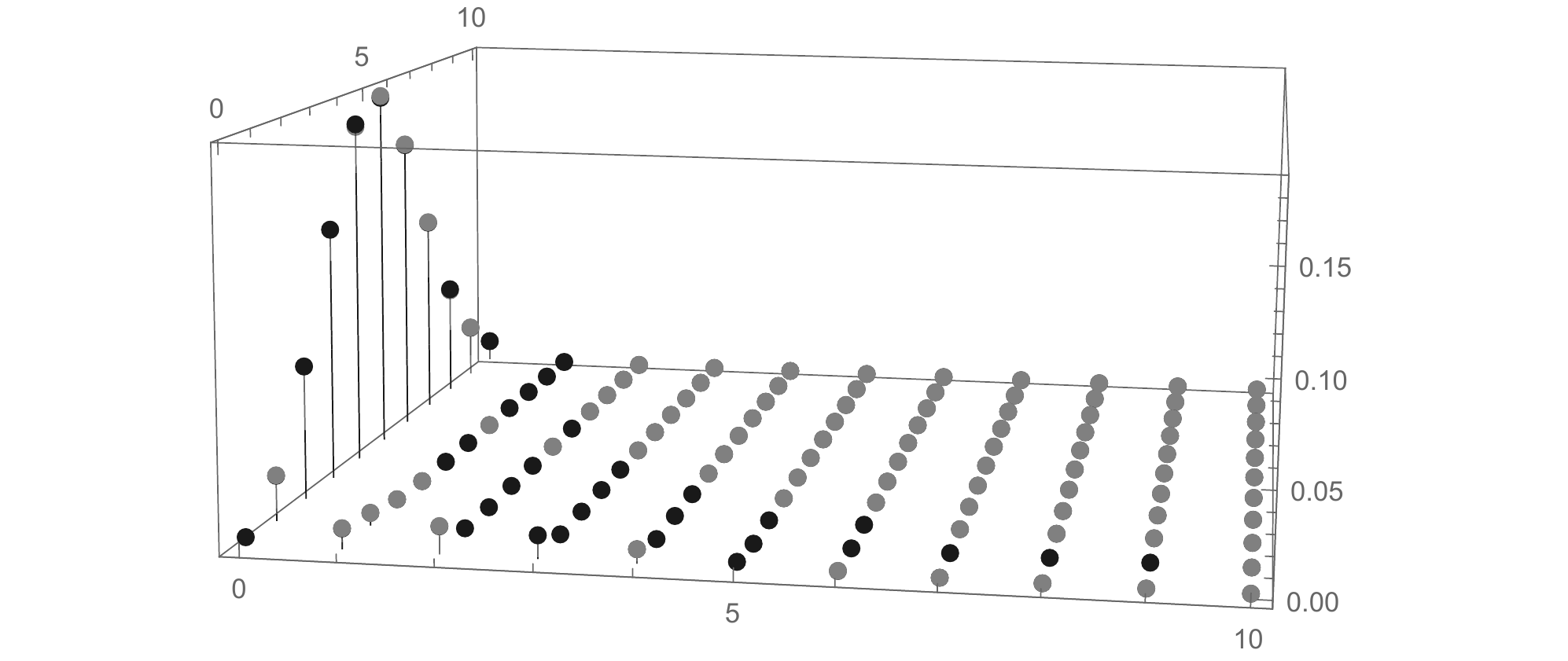} 
		\end{tabular} 
		\caption{p.m.f.s of the empirical (in gray) and of the BBCD distributions (in black), for (i) a sample of size 1000 and $n = 10$, $p_1 = 0.5$, $p_2 = 0.9$, $t = 0.8$, and for (ii) a sample of size 100000 and $n = 25$, $p_1 = 0.1$, $p_2 = 0.2$, $t = 0.1$}
		\label{histgibbs}
	\end{figure}
\end{center}

In Table \ref{tabgibbs}, we assess the performance of Gibbs sampler method to generate samples from the BBCD distribution in two scenarios and also the precision of the maximum likelihood estimates for the parameters $n_1$, $n_2$, $ p_1 $, $p_2$ and $t$. For a matter of simplicity and to reduce the computational time of the maximum likelihood estimations we consider $n_1=n_2=n$. From Table  \ref{tabgibbs} it is possible to observe that, in the first scenario, $n = 10$, $p_1 = 0.5$, $p_2 = 0.9$ and $t = 0.8$, for samples obtained using Gibbs sampling method of sizes big enough, for example $N\geq 1000$, it is possible to obtain reasonable maximum likelihood estimates of the parameters. In this first scenario we have $t=0.8$, close to 1, which points to a low correlation between the variables. In the second scenario, $n = 25$, $p_1 = 0.1$, $p_2 = 0.2$ and $t = 0.1$, the correlation is stronger since $t=0.1$ and close to 0, and bigger samples are needed in order to obtain a fair agreement between the exact and estimated values of the parameters. In Figure \ref{histgibbs}, we present the p.m.f.s of the empirical (in gray) and fitted BBCD (in black) distributions, in the two scenarios considered in Table \ref{tabgibbs}, taking $N=1000$ for the first scenario and $N=100000$ for the second scenario. This Figure shows the good fit between the empirical and the fitted BBCD distributions.
  
\setlength{\tabcolsep}{5pt}
\begin{table}[h!]
	\caption{Maximum likelihood estimates of $n$, $ p_1 $, $p_2$ and $t$ for samples of sizes $N$ from a BBCD distribution using Gibbs sampler method}
	\begin{center}
		\begin{tabular}{cccc|rcccc}
		$n$ & $p_1$ & $p_2$ & $t$ & $N$ &	$\hat{n}$ & $ \hat{p}_1 $ & $\hat{p}_2$ & $\hat{t} $  \\\hline
		 10 & 0.5 & 0.9 & 0.8 & 100   &	10 & 0.37303 & 0.88436 & 0.83132   \\
		    &     &     &     & 250   &	10 & 0.40986 & 0.89467 & 0.84159   \\
		    &     &     &     & 500   &	10 & 0.53465 & 0.89819 & 0.79151   \\
		    &     &     &     & 1000  &	10 & 0.47977 & 0.89905 & 0.80359   \\
		    &     &     &     & 2500  &	10 & 0.49307 & 0.89952 & 0.80356   \\
		    &     &     &     & 5000  &	10 & 0.49949 & 0.89978 & 0.79913   \\\hline
		 25 & 0.1 & 0.2 & 0.1 & 100   &	23 & 0.09688 & 0.23313 & 0.30804   \\
		    &     &     &     & 250   &	29 & 0.08049 & 0.17482 & 0.22896   \\
		    &     &     &     & 500   &	23 & 0.07646 & 0.22170 & 0.18795   \\
		    &     &     &     & 1000  &	29 & 0.08789 & 0.17109 & 0.15450   \\
		    &     &     &     & 2500  &	29 & 0.08449 & 0.17278 & 0.11737   \\
		    &     &     &     & 5000  &	26 & 0.10164 & 0.19089 & 0.10958   \\
   		    &     &     &     & 10000 &	26 & 0.10164 & 0.19089 & 0.10958   \\
   		    &     &     &     & 50000 &	24 & 0.10345 & 0.20834 & 0.10305   \\
   		    &     &     &     & 100000&	25 & 0.09989 & 0.20001 & 0.09996   \\\hline
		\end{tabular}
		\vspace{-.6cm}
	\end{center}
	\label{tabgibbs}
\end{table}

\section{Real data application}

\begin{center}
	\begin{figure}[h]
		\includegraphics[scale = 0.5]{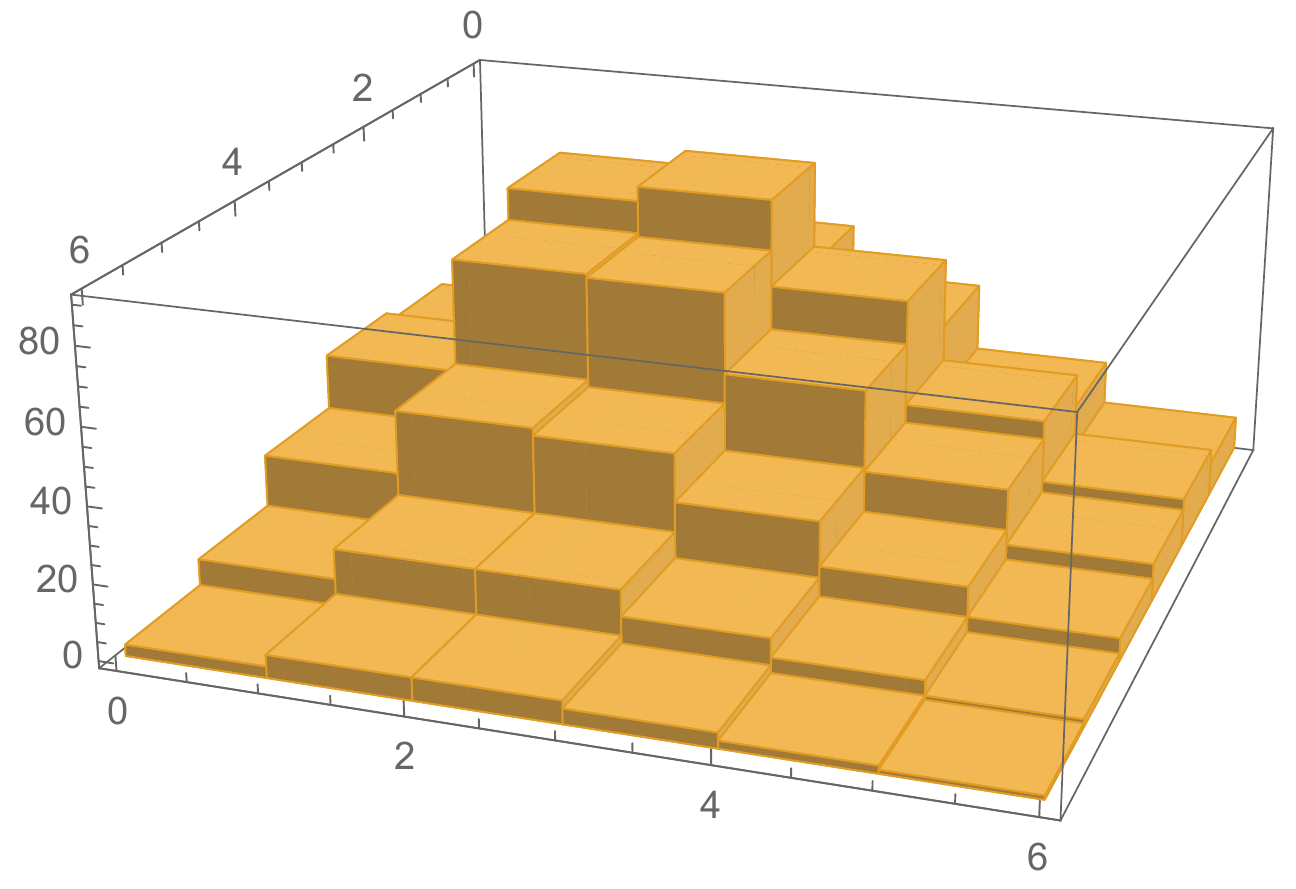} 
		\caption{Histogram of the data of seeds and plants grown}
		\label{figseeds}
	\end{figure}
\end{center}

We consider the data about seeds and plants grown in Rao (1990) and also in Table 1 in Lakshminarayana et al. (1999) and also in Ghosh et al. (2020). The data set reports the number of seeds and plants grown over a plot of size five square feet. In Table \ref{desc} descriptive measures of the observed data is presented while   in Figure \ref{figseeds} we present the histogram of the data.

\setlength{\tabcolsep}{3pt}
\begin{table}[h]
	\caption{Descriptive measures of the data of seeds and plants grown}
	\begin{center}
		\begin{tabular}{c|cccccc|}
				  &	$\min$ & $Q_{0.25}$ & Median & Mean & $Q_{0.75}$ & $\max$  \\\hline
			$X_1$ &    0   &  1.000     & 2.000  & 1.692&  2.000     & 5.000   \\\hline
			$X_2$ &    0   &  1.000     & 2.000  & 2.013&  3.000     & 5.000   \\\hline
			\multicolumn{1}{c|}{Correlation} & \multicolumn{6}{c|}{-0.0938} \\\hline
		\end{tabular}
	\end{center}
	\label{desc}
\end{table} 

In Table \ref{seeds}, we present the estimated values of $p_1$, $p_2$ and $t$, for different choices of $n_1=n_2=n$ together with the sample and population correlations and with the $p$-value of $\chi^{2}$ goodness-of-fit test for the BBCD distribution. In this table the row in text bold presents the values obtained by maximum likelihood estimation for all the parameters  $n$, $ p_1 $, $p_2$ and $t$. In rest of the rows, the ones that are not in bold, we assume a specific value for $n$ and only the parameters $ p_1 $, $p_2$ and $t$ are estimated.

\setlength{\tabcolsep}{5pt}
\begin{table}[h!]
	\caption{Analysis of the fit of the BBCD distribution to the data of seeds and plants grown}
	\begin{center}
		\begin{tabular}{ccccccc}
			             &               &             &            & \multicolumn{2}{c}{Correlation}& \\\hline
			         $\hat{n}$ & $ \hat{p}_1 $ & $\hat{p}_2$ & $\hat{t} $ &  BBCD  &  data  & $p$-value \\\hline
			           8  &  0.238  &  0.277  &  0.926  &  -0.107  &  -0.094  &  0.001 \\\hline
			          10  &  0.189  &  0.220  &  0.934  &  -0.101  &  -0.094  &  0.363  \\\hline
		    \textbf{14}   & \textbf{ 0.134}  &  \textbf{0.156}  &  \textbf{0.943}  &  \textbf{-0.092}  &  \textbf{-0.094}  &  \textbf{0.788} \\\hline
			          15  &  0.125  &  0.146  &  0.942  &  -0.0943  &  -0.094  &  0.852 \\\hline
				      20  &  0.0935  &  0.109  &  0.946  &  -0.0911  &  -0.094  &  0.931	\\\hline
				      30  &  0.0621  &  0.0727  &  0.949  &  -0.0881  &  -0.094  &  0.987   \\\hline
				      40  &  0.0465  &  0.0545  &  0.951  &  -0.0867  &  -0.094  &  0.989 \\\hline
				      50  &  0.0372  &  0.0435  &  0.952  &  -0.0859  &  -0.094  &  0.989 \\\hline
				     100  &  0.0186  &  0.0217  &  0.954  &  -0.0842  &  -0.094  &  0.978 \\\hline
		\end{tabular}
		\vspace{-.6cm}
	\end{center}
	\label{seeds}
\end{table} 

\begin{center}
	\begin{figure}[h]
		\includegraphics[scale = 0.5]{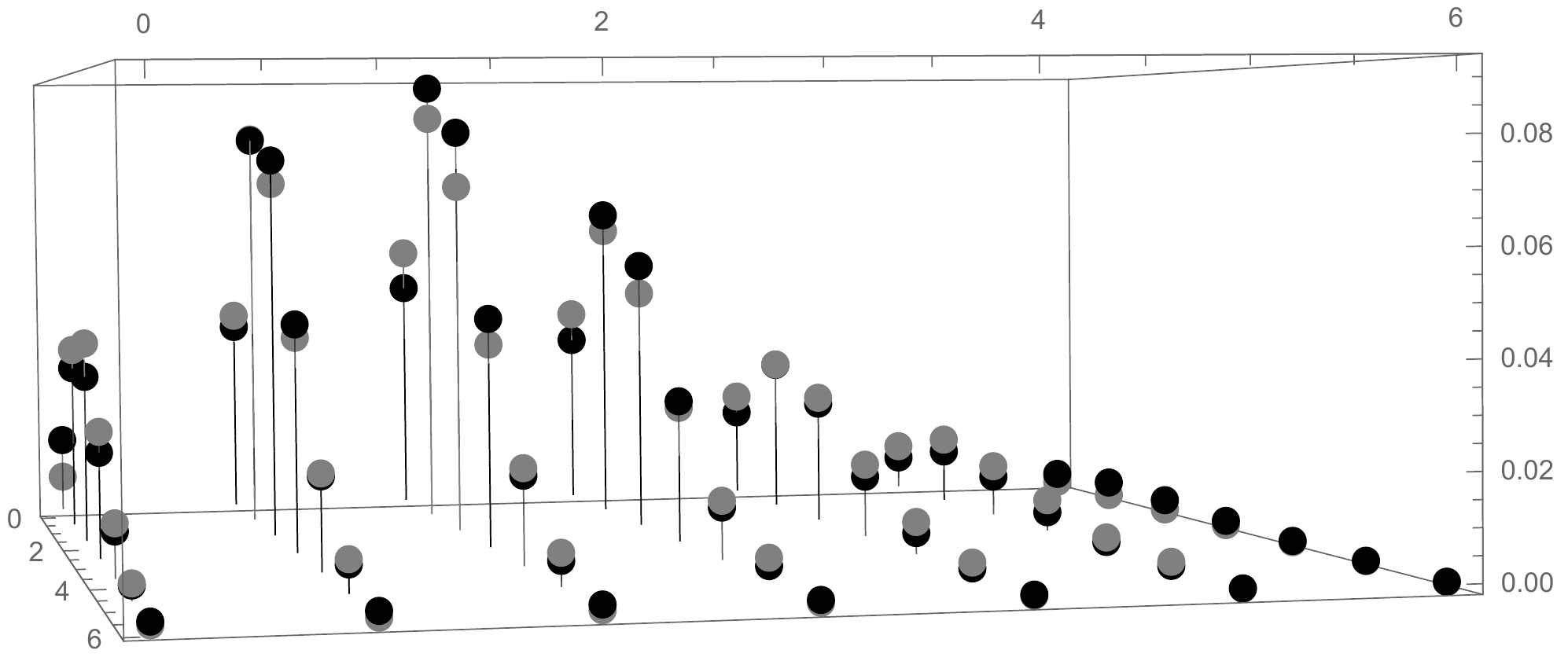} 
		\caption{p.m.f.s of the empirical (in gray) and of the BBCD distributions (in black)}
		\label{figfit}
	\end{figure}
\end{center}

We may observe that for $n\geq 14$ the sample and population correlations are similar  and also that the $p$-value of the $\chi^{2}$ goodness-of-fit test for the BBCD distribution is close to 1, suggesting an excellent fit of the BBCD distribution to the data. It is also interesting to note that when $n$ increases the values of $n_1\times p_1$, $n_2\times p_2$ and $t$ converge to the values obtained in Ghosh et al. (2020) for the same application and, respectively, for the parameters $\lambda_1$, $\lambda_2$ and $\lambda_3$ of the bivariate Poisson distribution  as established in Theorem 9.  In Figure \ref{figfit}, the p.m.f.s of the empirical and of the fitted BBCD distributions are presented for the case $n=14$ supporting the good fit of the BBCD distribution to the data.

\section{Conclusion:}
 In this paper, we have studied a  bivariate binomial distribution via conditional specification originally proposed by Arnold et al. (1991, 1999).
The model is useful for bivariate-dependent count data when negative  correlation
structure is observed. The flexibility and the importance of the model are discussed.
One real data example is provided to illustrate the importance of the proposed model.
This example deals with  negative correlation and over-dispersed marginals.
 It is envisaged  that the BBCD studied here will be a viable alternative to the existing bivariate count data dealing with the kind of data
sets considered here in various other real life scenarios. A multivariate extension of the BBCD will be explored in a separate article. However, appropriate real life scenarios must be found for its possible application albeit computational complexity that is expected for higher dimensions.\\


\begin{thebibliography}{99}

\bibitem{as}
Arnold, B. C., \& Strauss, D. J. (1991). Bivariate distributions with conditionals in prescribed exponential families. \textit{Journal of the Royal Statistical Society}: Series B, 53(2), 365-375.

\bibitem{Acs}
Arnold, B.C.,  Castillo, E.,  \& Sarabia, J.M. (1999). \textit{Conditional Specification of Statistical Models.} Springer, New York.

\bibitem{Ag}
Aitken, A.C., \& Gonin, H.T. (1935). On fourfold sampling with and without replacement. \textit{Proceedings of the Royal Society of Edinburgh.} 55, 114-125.



\bibitem{bu}
Becker, N. G., \& Utev, S. (2002). Multivariate discrete distributions with a product-type dependence. \textit{Journal of multivariate analysis}, 83(2), 509-524.

\bibitem{bh}
Biswas, A., \& Hwang, J. S. (2002). A new bivariate binomial distribution. \textit{Statistics \& probability letters}, 60(2), 231-240.

\bibitem{cp}
Consul, P. C. (1974). A simple urn model dependent upon predetermined strategy. \textit{Sankhya: The Indian Journal of Statistics}, Series B, 391-399.

\bibitem{cs}
Crowder, M., \& Sweeting, T. (1989). Bayesian inference for a bivariate binomial distribution. \textit{Biometrika}, 76(3), 599-603.

\bibitem{ga}
Gelfand, A. E. (2000). Gibbs sampling. Journal of the American statistical Association, 95(452), 1300-1304.



\bibitem{Ghosh}
Ghosh, I., Marques, F., \& Chakraborty, S. (2020) A new bivariate Poisson distribution via conditional specification: properties and applications. Journal of Applied Statistics, DOI: 10.1080/02664763.2020.1793307.


\bibitem{hj}
Hamdan, M. A., \& Jensen, D. R. (1976). A bivariate binomial distribution and some applications. \textit{Australian Journal of Statistics}, 18(3), 163-169.

\bibitem{ht}
Hamdan, M. A., \& Tsokos, C. P. (1971). A model for physical and biological problems: the bivariate-compounded Poisson distribution. \textit{Revue de l'Institut International de Statistique}, 60-63.

\bibitem{hw}
Harkness, W. L. (1965). Properties of the extended hypergeometric distribution. The Annals of Mathematical Statistics, 36(3), 938-945.

\bibitem{Jkb}
Johnson, N.L., Kotz, S., \& Balakrishnan, N. (1997). \textit{Discrete Multivariate Distributions.} John Wiley \& Sons, New York.


\bibitem{Kk}
Kocherlakota, S., \& Kocherlakota, K. (1992). \textit{Bivariate Discrete Distributions.} New York, Marcel
Dekker.


\bibitem{Lak}
Lakshminarayana, J., Pandit, S. N. N., \& Srinivasa Rao, K. (1999). On a bivariate Poisson distribution. \textit{Communications in Statistics--Theory and Methods}, 28(2), 267-276.


\bibitem{lc}
Le, C. T. (1984). A symmetric bivariate binomial distribution and its application to the analysis of clustered samples in medical research. \textit{Biometrical journal}, 26(3), 289-294.

\bibitem{Lee}
Lee, H., Cha, J. H., \& Pulcini, G. ( 2017). Modeling Discrete Bivariate Data with Applications to Failure and Count Data. \textit{Quality and Reliability Engineering International} 33, 1455-1473.

\bibitem{lt}
Ling, K. D., \& Tai, T. H. (1989). On bivariate binomial distributions of order k. \textit{Soochow Journal of Mathematics}, 16, 211-220.

\bibitem{Loukas}
Loukas, S., \&  Kemp, C. (1986). On the Chi-Square Goodness-of-Fit Statistic for Bivariate Discrete Distributions. \textit{Journal of the Royal Statistical Society. Series D (The Statistician)}, 35(5), 525-529.

\bibitem{ms}
Mishra, A., \& Singh, S. K. (1996). Moments of a quasi-binomial distribution. \textit{Progress of Mathematics}, 30, 59-67.

\bibitem{Ne}
Nelsen, R. B. (2006).  \textit{An Introduction to Copulas}. 2nd edition. New York, Springer.

\bibitem{Nika}
Nikoloulopoulos, A. K. (2013a). Copula-based models for multivariate discrete response data. \textit{In P. Jaworski, F. Durante, and W. Hardle, editors, Copulae in Mathematical and Quantitative Finance}, 231-249.

\bibitem{Nikb}
Nikoloulopoulos, A. K. (2013b). On the estimation of normal copula discrete regression models using the continuous extension and simulated likelihood. \textit{In P. Jaworski, F. Durante, and W. Hardle, editors, Copulae in Mathematical and Quantitative Finance}, 231-249.
A. K. Nikoloulopoulos. \textit{Journal of Statistical Planning and Inference}, 143(11):1923-1937.



\bibitem{}
Ong, S.H., \& Ng, C.M. (2013). A bivariate generalization of the non-central
negative binomial distribution. \textit{Communications in Statistics - Simulation and
Computation}, 42, 570-585.

\bibitem{pd}
Papageorgiou, H., \& David, K. M. (1994). On countable mixtures of bivariate binomial distributions. \textit{Biometrical journal}, 36(5), 581-601.

\bibitem{RaoS}
Rao, S. (1990). Experimental studies on the yield of groundnuts in coastal
region. \textit{Technical Report,} Andhra University, Visakhapatnam.

\bibitem{}
Sun, K., \& Basu, A. P. (1995). A characterization of a bivariate binomial distribution. \textit{Statistics \& probability letters}, 23(4), 307-311.

\bibitem{tt}
Takeuchi, K., \& Takemura, A. (1987). On sum of $0-1$ random variables I. Univariate case. \textit{Annals of the Institute of Statistical Mathematics}, 39(1), 85-102.

\end{thebibliography}
\end{document}